\documentclass[twocolumn,apjl]{emulateapj}

\shorttitle{DARK MATTER HALOS AND GALAXY FORMATION MODELS. I.}
\shortauthors{SEIGAR ET AL.}

\begin{document}


\title{Constraining dark matter halo profiles and galaxy formation models using spiral arm morphology. I. Method outline}


\author{Marc S. Seigar\altaffilmark{1}, James S. Bullock\altaffilmark{1}, Aaron J. Barth\altaffilmark{1} and Luis C. Ho\altaffilmark{2}}
\altaffiltext{1}{University of California, Irvine, Department of Physics \& Astronomy, 4129 Frederick Reines Hall, Irvine, CA 92697-4575}
\altaffiltext{2}{The Observatories of the Carnegie Institution of Washington, 813 Santa Barbara Street, Pasadena, CA 91101}



\begin{abstract}
We investigate the use of spiral arm pitch angles 
as a probe of disk galaxy mass profiles.
We confirm our previous result that 
spiral arm pitch angles (P) are well-correlated with the rate of shear
(S) in disk galaxy rotation curves, by using a much larger sample
(51 galaxies) than used previously (17 galaxies). We use this 
correlation to argue that
imaging data alone can provide a powerful probe
of galactic mass distributions out to large lookback times.
  In contrast to previous work,
we show that observed spiral
arm pitch angles are similar when measured in the optical (at 0.4 $\mu$m)
and the near-infrared (at 2.1 $\mu$m) with a mean difference of 
$2\fdg3\pm2\fdg7$.
This is then used to
strengthen the known correlation between P and S using $B$ band images.
We then use two example galaxies to
demonstrate how an inferred shear rate coupled with
a bulge-disk decomposition model and a Tully-Fisher derived
velocity normalization can be used to place constraints
on a galaxy's baryon fraction and dark matter halo profile.
We show that ESO 582-G12, a galaxy with a high shear rate
(slightly declining rotation curve)
at $\sim 10$ kpc, favors an adiabatically contracted halo, with
high initial NFW concentration ($c_{\rm vir} > 16$) and 
a high fraction of halo baryons in the form of stars ($\sim 15-40$\%).
In contrast, IC 2522 has a low shear rate (rising rotation curve)
at $\sim 10$ kpc and favors non-adiabatically contracted models
with low NFW concentrations ($c_{\rm vir} \simeq 2-8$) and 
a low stellar baryon fraction $< 10$\%.

\end{abstract}


%% subject header.  Objects should be in the appropriate "individual"
%% headers (e.g. quasars: individual, stars: individual, etc.) with the
%% additional provision that the total number of headers, including each
%% individual object, not exceed six.  The \objectname{} macro, and its
%% alias \object{}, is used to mark each object.  The macro takes the object
%% name as its primary argument.  This name will appear in the paper
%% and serve as the link's anchor in the electronic edition if the name
%% is recognized by the data centers.  The macro also takes an optional
%% argument in parentheses in cases where the data center identification
%% differs from what is to be printed in the paper.



\section{Introduction}

The correlation found between spiral arm pitch angle and rotation
curve shear rate (Seigar, Block \& Puerari 2004; Seigar et al.\ 2005) 
suggests that there is a link between the tightness of spiral structure
and the central mass concentration in spiral galaxies. The shear rate, $S$,
a dimensionless quantity, 
can be measured directly from rotation curves and is defined as follows,
\begin{equation}
\label{shearrate}
S=\frac{A}{\omega}=\frac{1}{2}\left(1-\frac{R}{V}\frac{dV}{dR}\right),
\end{equation}
where A is the first Oort constant, $\omega$ is the angular velocity and
$V$ is the velocity at a radius $R$. The shear rate depends upon the
shape of the rotation curve. For a rotation curve that remains flat $S=0.5$,
for a falling rotation curve $S>0.5$ and for a continually rising rotation
curve $S<0.5$. As the shape of a rotation curve depends on the mass
distribution, the shear rate at any given position depends upon the mass
within that radius, or the central mass concentration. As a result, the
spiral arm pitch angle is dependent upon the central mass concentration, and 
this is consistent with the expectations of most spiral density wave models
(e.g. Bertin et al.\ 1989a, b; Bertin 1991, 1993, 1996; Bertin \& Lin 1996; 
Fuchs 1991, 2000), although density wave models predict that pitch angles
also depend on stability (i.e. the Toomre $Q$-parameter).

The modal theory of spiral structure (Bertin et al.\ 1989a, b; Bertin 1991, 
1993, 1996; Bertin \& Lin 1996) predicts that the tightness of the arms
comes from the central mass concentration. Galaxies with higher central mass 
concentrations, i.e. higher overall densities (including dark matter), are
predicted to have more tightly wound spiral arms. The models of Fuchs (1991,
2000) result in disks with rigidly rotating spiral modes, wherein bulges act
as inner reflectors of waves or modes induced by the swing amplification
method, thus leading to modal spiral waves, which form as a result of Toomre
swing amplification (Toomre 1981). Fuchs (2000) adopts a stellar-dynamical
analogue (Julian \& Toomre 1966)
of the Goldreich \& Lynden-Bell (1965) sheet, which describes the
local dynamics of an unbounded patch of thin, 
differentially rotating stellar disk. He then increases $Q$ in the Goldreich 
\& Lynden-Bell (1965) sheet, which has the affect that the bulge acts as
an inner boundary. The
result is that, instead of shearing density waves (as in the unbounded sheet),
spiral modes appear. These models best show how central mass concentration
correlates with spiral arm pitch angle. If the disk is very light (low
$\sigma$ where $\sigma$ is the disk density) the mode can be very tight, and
we are in the domain of small epicycles. Formally, if the stability parameter
$Q=c\kappa / \pi G \sigma$ is close to unity, the value of $c$ must also 
be small, where $c$ is the radial velocity dispersion and $\kappa$ is the 
epicyclic frequency. If we increase the relative mass of the disk, we find a trend
towards more open structures.

Recent observational studies of spiral structure have highlighted a 
difference in the morphologies seen in the optical compared with the
near-infrared. The classification of galaxies by Hubble type 
(Hubble 1926) is performed in the optical regime, where dust extinction
still has a large effect, and where the light is dominated by the
young Population I stars. In the near-infrared dust extinction is minimized
and the light is dominated by old and intermediate age stars (Rhoads 1998;
Worthey 1994). Previous work has found that optically determined Hubble type
is not correlated with the near-infrared morphology. There is not a good
correlation between Hubble type and near-infrared bulge-to-disk ratio or
near-infrared spiral arm pitch angle (de Jong 1996; Seigar \& James 1998a,
b). However, even in the optical regime, the correlation between 
quantitative pitch angle and Hubble type is weak (Kennicutt 1981), even
though the tightness of the spiral arm pattern, as judged by eye, is one of
the defining Hubble morphological criteria. Furthermore, it has been 
shown that near-infrared morphologies of some spiral galaxies can be 
significantly different from their optical morphologies (Block \& 
Wainscoat 1991; Block et al.\ 1994; Thornley 1996; Seigar \& James 1998a, b;
Seigar, Chorney \& James 2003). Galaxies with flocculent spiral structure
in the optical may present grand-design spiral structure in the near-infrared
(Thornley 1996; Seigar et al.\ 2003). These results suggest that the optical
morphology gives incomplete information about the nature of spiral structure
in disk galaxies. This is the basis of the near-infared studies of
Block et al.\ (1999) and Seigar et al.\ (2005).

However, work by Eskridge et al.\ (2002) showed that 
when assigning a Hubble type to galaxies in several different wavebands,
from the optical to the near-infrared, a correlation exists between the
optically determined Hubble type and that determined in the near-infrared.
This suggests that, while there may be large small-scale differences 
between optical and near-infrared morphologies of spiral structure, the 
large-scale differences are minimal. As a result, it may be possible that
there is no significant difference between the large-scale spiral arm
pitch angle, regardless of whether it is 
determined in the optical or near-infrared for the regular Grand-Design
spirals investigated here. It should be noted that for flocculent spiral
structure it is difficult to determine a pitch angle for the two-armed
component in the optical, while a $m=2$ Grand-Design structure may be
seen in the near-infrared (e.g.\ Thornley 1996; Grosbol \& Patsis 1998;
Seigar, Chorney \& James 2003). As
a result it may be possible to use optical imaging to expand the sample
of galaxies having quantitative measurements of spiral arm pitch angle
and therefore strengthen the
correlation between spiral arm pitch angle and rotation curve shear rate
reported by Seigar et al.\ (2005).

In addition  to providing an important constraint  on models of spiral
arm formation,  the correlation  between  spiral arm pitch  angle  and
shear rate  opens  up a  fundamentally new  approach for probing  mass
distributuions  in spiral galaxies.  This   approach relies on imaging
data alone without the    need for full rotation    curve information.
Specifically, the  Tully-Fisher relation (Tully \& Fisher 1977)
for spiral  galaxies coupled
with the shear rate - pitch angle relation can be  used to determine a
rotation  curve normalization  {\it and}   slope.   In this  paper  we
explicitly demonstrate how, given  a bulge-disk decomposition, a  pitch
angle determination, and a Tully-Fisher normalizaion, one can constrain
galaxy mass distributions,    dark matter halo  concentrations,   and
other galaxy  formation  parameters.  We  obtain our constraints within  the
context of the  the standard framework  of disk formation put forth by
Fall \& Efstathiou (1980) and Blumenthal et al.\ (1986).
  
In principle, the technique we demonstrate here
can be applied generally to a large sample of galaxies and to galaxies
at high redshift, when spiral arms are detected.
This is the first in a series of papers in which we use spiral
arm pitch angles to determine the mass distribution in spiral galaxies.
In this paper we determine mass concentrations in a few disk galaxies
from their spiral arm pitch angles (or shear rates) and their disk masses
and scalelengths (determined via a bulge-disk decomposition technique).
In future papers we will apply the outline methods 
in this paper to a large sample
of nearby disk galaxies and to a large sample of galaxies at
higher redshift in order to determine the evolution in the central mass
concentrations of disk galaxies as a function of look-back time.

This paper is arranged as follows. In section 2 we describe the data used
and the methods for determining spiral arm pitch angles and rotation
curve shear rates. In section 3 we compare pitch angles in the near-infrared
and optical. In section 4 we discuss the relationship between spiral arm
pitch angle and rotation curve shear rate found by Seigar et al.\ (2005) and
we use more data to strengthen this correlation. In section 5 we use
a bulge-disk decomposition routine and
an adiabatic infall model for disk galaxy formation to investigate the link 
between shear rate and central mass concentration and we summarize our
findings in section 6.  
%Throughout this work we assume a Hubble
%constant of $h=0.75$ in units of $100$ km s$^{-1]$ Mpc^{-1} and a
%cosmology with $\Omega_m = 1 - \Omega_{\Lambda} = 0.3$.

\section{Observations and data reduction}

%% In a manner similar to \objectname authors can provide links to dataset
%% hosted at participating data centers via the \dataset{} command.  The
%% second curly bracket argument is printed in the text while the first
%% parentheses argument serves as the valid data set identifier.  Large
%% lists of data set are best provided in a table (see Table 3 for an example).
%% Valid data set identifiers should be obtained from the data center that
%% is currently hosting the data.

We have made use of $H$ and $B$ band images of 57 galaxies from the Ohio 
State University Bright Spiral Galaxy Survey (OSUBSGS; e.g. Eskridge et al.\ 
2002). These galaxies were chosen to be as face-on as possible (with a 
ratio of the minor- to major-axis $b/a>0.5$) so that
a comparison of spiral arm morphology could be made in the optical and
near-infrared.

We also include $B$ band images of 31 galaxies that also have measured
rotation curves (Mathewson et al.\ 1992; Persic \& Salucci 1995). We
observed these galaxies as part of an ongoing 
survey of the 600 brightest southern
hemisphere galaxies. The observations make use
of the CCD direct imaging camera at the 2.5-m du Pont telescope at the Las
Campanas Observatory in Chile. Integration times used were $2\times 360$ 
seconds. For this work, we selected from the full imaging dataset a sample
of galaxies having a ratio of the minor- to major-axis
$b/a>1/3$, so that spiral arm morphology can be studied, and rotation 
curves are also available. Of these galaxies, 6 also have $K_s$ band data.
The $K_s$ band images were observed using the Wide-field Infrared Camera
(WIRC), also at the 2.5-m du Pont telescope. A 5-point dither pattern
was used with integration times of $3\times60$ seconds in each position.
The total exposure was therefore 900 seconds.

We also include $B$ and $K_s$ band images of a further 3 galaxies from our
survey, which do not have rotation curves. The selection for these galaxies
satisfies the same selection criteria that was used to select the OSUBSGS
galaxies above.

\subsection{Measurement of spiral arm pitch angles}

\begin{deluxetable*}{llcccll}
\tablecolumns{7}
\tablewidth{7in}
\tablecaption{$B$ band and $H$ band spiral arm pitch angles for 57 galaxies from the
OSUBSGS.}
\label{tab1}
\tablehead{
\colhead{Galaxy}& \colhead{Hubble}& \colhead{b/a}         & \colhead{PA}       & \colhead{Radial Ranges}             & \colhead{$P_B$}          & \colhead{$P_H$}          \\
\colhead{Name}  & \colhead{Type}  &                       & \colhead{}         & \colhead{}                          & \colhead{}               & \colhead{}\\
\colhead{}                & \colhead{}                     & \colhead{}            & \colhead{(degrees)}& \colhead{(arcsec)}                  & \colhead{(degrees)}      & \colhead{(degrees)}      \\
}
\startdata
NGC 150         & SBbc            & 0.50        &  118           & 70--126; 65--131; 75--121    & 8.4$\pm$0.1  & 9.4$\pm$0.8  \\
NGC 157         & SABbc           & 0.64        &  30            & 65--121; 60--126; 70--116    & 25.2$\pm$0.7 & 25.8$\pm$1.1 \\
NGC 289         & SABbc           & 0.71        &  130           & 40--150; 35--155; 45--145    & 25.7$\pm$0.9 & 18.4$\pm$2.2 \\
NGC 578         & SABc            & 0.63        &  110           & 65--125; 60--130; 70--120    & 18.0$\pm$0.2 & 18.8$\pm$0.5 \\
NGC 613         & SBbc            & 0.76        &  120           & 73--135; 68--140; 78--130    & 26.4$\pm$1.0 & 41.8$\pm$1.8 \\
NGC 864         & SABc            & 0.74        &  20            & 55--143; 50--148; 60--138    & 46.4$\pm$1.3 & 42.9$\pm$3.0 \\
NGC 1073        & SBc             & 0.92        &  15            & 73--183; 68--188; 78--178    & 17.6$\pm$0.8 & 11.9$\pm$0.8 \\
NGC 1087        & SABc            & 0.59        &  5             & 57--147; 52--152; 62--142    & 31.1$\pm$1.6 & 32.7$\pm$0.5 \\
NGC 1187        & SBc             & 0.75        &  130           & 90--132; 85--137; 95--127    & 15.7$\pm$0.9 & 15.1$\pm$1.5 \\
NGC 1241        & SBb             & 0.61        &  140           & 25--51; 20--56; 30--46       & 50.3$\pm$1.4 & 44.2$\pm$3.0 \\
NGC 1300        & SBbc            & 0.66        &  106           & 80--116; 75--121; 85--111    & 31.7$\pm$1.1 & 29.0$\pm$1.4 \\
NGC 1350        & SBab            & 0.54        &  0             & 75--87; 72--90; 78--84       & 48.8$\pm$4.8 & 52.0$\pm$3.8 \\
NGC 1792        & SAbc            & 0.50        &  137           & 45--211; 40--216; 50--206    & 42.1$\pm$4.3 & 47.0$\pm$1.8 \\
NGC 2090        & SAb             & 0.50        &  13            & 64--272; 59--277; 69--267    & 20.3$\pm$0.3 & 20.3$\pm$0.8 \\
NGC 2139        & SABcd           & 0.73        &  140           & 50--118; 45--123; 55--113    & 26.4$\pm$3.3 & 30.2$\pm$1.9 \\
NGC 2964        & SABbc           & 0.55        &  97            & 70--162; 65-167; 75--157     & 46.9$\pm$3.9 & 47.5$\pm$2.5 \\
NGC 3223        & SAbc            & 0.61        &  135           & 32--74; 27--79; 37--69       & 10.7$\pm$2.0 & 7.1$\pm$1.4  \\
NGC 3261        & SBbc            & 0.76        &  85            & 30--90; 25--95; 35--86       & 21.2$\pm$1.7 & 22.4$\pm$1.1 \\
NGC 3338        & SAc             & 0.61        &  100           & 74--164; 69--169; 79--159    & 13.6$\pm$0.3 & 16.2$\pm$0.6 \\
NGC 3507        & SBb             & 0.85        &  110           & 120--196; 115--201; 125--191 & 24.0$\pm$2.4 & 23.2$\pm$2.0 \\
NGC 3513        & SBc             & 0.79        &  75            & 80--130; 75--135; 85--125    & 28.1$\pm$4.3 & 25.2$\pm$2.9 \\
NGC 3583        & SBb             & 0.64        &  125           & 60--84; 57--87; 63--81       & 32.5$\pm$2.6 & 36.9$\pm$1.9 \\
NGC 3646        & SABc            & 0.56        &  50            & 20--52; 15--57; 25--47       & 21.7$\pm$3.0 & 17.0$\pm$2.3 \\
NGC 3686        & SBbc            & 0.78        &  15            & 99--169; 94-174; 104--164    & 15.3$\pm$4.2 & 13.4$\pm$2.8 \\
NGC 3726        & SABc            & 0.69        &  10            & 100--258; 95--263; 105--253  & 32.0$\pm$4.4 & 31.7$\pm$1.2 \\
NGC 3810        & SAc             & 0.70        &  15            & 112--200; 107--205; 117--195 & 31.0$\pm$2.8 & 30.4$\pm$4.3 \\
NGC 3887        & SBbc            & 0.76        &  20            & 93--163; 88--168; 98--158    & 24.9$\pm$3.9 & 24.4$\pm$2.6 \\
NGC 3893        & SABc            & 0.62        &  165           & 102--218; 97--223; 107--213  & 18.2$\pm$0.5 & 19.5$\pm$0.5 \\
NGC 4027        & SBdm            & 0.75        &  167           & 60--126; 55--131; 65--121    & 35.7$\pm$5.0 & 37.6$\pm$3.2 \\
NGC 4030        & SAbc            & 0.71        &  27            & 51--161; 46--166; 56--156    & 19.8$\pm$3.2 & 20.6$\pm$3.5 \\
NGC 4051        & SABbc           & 0.75        &  135           & 185--257; 180--262; 190--267 & 18.3$\pm$3.6 & 17.7$\pm$1.7 \\
NGC 4145        & SABd            & 0.73        &  100           & 93--211; 88--216; 98--206    & 43.3$\pm$2.9 & 43.6$\pm$2.7 \\
NGC 4414        & SAc             & 0.56        &  155           & 142--290; 137--295; 147--285 & 13.2$\pm$2.6 & 7.7$\pm$1.8  \\
NGC 4548        & SBb             & 0.80        &  150           & 250--386; 245--391; 255--381 & 24.8$\pm$0.4 & 31.0$\pm$0.6 \\
NGC 4580        & SABa            & 0.76        &  165           & 82--218; 77--223; 87--213    & 22.1$\pm$4.8 & 22.1$\pm$2.4 \\
NGC 4654        & SABcd           & 0.57        &  128           & 99--199; 94--204; 104-194    & 32.1$\pm$2.6 & 30.4$\pm$1.4 \\
NGC 4930        & SBbc            & 0.82        &  40            & 43--77; 38--82; 48--72       & 40.4$\pm$4.2 & 41.2$\pm$3.9 \\
NGC 4939        & SAbc            & 0.51        &  10            & 24--76; 19--81; 29--71       & 22.4$\pm$1.8 & 22.3$\pm$1.7 \\
NGC 4995        & SABb            & 0.64        &  95            & 31--145; 26--150; 36--140    & 17.8$\pm$0.7 & 18.0$\pm$0.5 \\
NGC 5054        & SAbc            & 0.59        &  160           & 51--127; 46--132; 56--122    & 49.3$\pm$3.4 & 47.7$\pm$2.7 \\
NGC 5085        & SAc             & 0.88        &  38            & 20--138; 15--143; 25--133    & 16.3$\pm$2.8 & 16.9$\pm$1.3 \\
NGC 5247        & SAbc            & 0.88        &  20            & 47--181; 42--186; 52--176    & 49.5$\pm$0.7 & 46.3$\pm$0.5 \\
NGC 5371        & SABbc           & 0.80        &  8             & 42--78; 37--83; 47--73       & 38.7$\pm$1.2 & 37.1$\pm$0.9 \\
NGC 5483        & SBc             & 0.92        &  25            & 50--124; 55--119; 45--129    & 29.6$\pm$1.3 & 31.0$\pm$1.2 \\
NGC 5921        & SBbc            & 0.82        &  130           & 70--140; 65--145; 75--135    & 25.1$\pm$1.0 & 26.6$\pm$0.8 \\
NGC 6215        & SAc             & 0.86        &  78            & 47--141; 42--146; 52--136    & 36.7$\pm$4.0 & 34.3$\pm$3.6 \\
NGC 6221        & SBbc            & 0.71        &  5             & 73--135; 68--140; 78--130    & 40.4$\pm$3.2 & 32.8$\pm$1.3 \\
NGC 6300        & SBb             & 0.67        &  118           & 99--179; 94--184; 104--174   & 26.2$\pm$2.9 & 26.4$\pm$1.1 \\
NGC 6384        & SABbc           & 0.66        &  30            & 60--126; 55--131; 65--121    & 17.8$\pm$1.2 & 17.0$\pm$0.8 \\
NGC 6907        & SBbc            & 0.82        &  46            & 40--58; 38--60; 42--56       & 20.3$\pm$0.7 & 24.7$\pm$0.5 \\
NGC 7083        & SABc            & 0.59        &  5             & 32--78; 29--81; 35--75       & 24.9$\pm$0.9 & 25.7$\pm$0.5 \\
NGC 7412        & SABc            & 0.74        &  65            & 50--90; 45--95; 55-85        & 29.4$\pm$2.2 & 28.6$\pm$1.7 \\
NGC 7418        & SABcd           & 0.74        &  139           & 63--151; 58--156; 68--146    & 22.3$\pm$3.7 & 23.3$\pm$2.2 \\
NGC 7479        & SBc             & 0.76        &  25            & 48--82; 45--85; 51--79       & 17.8$\pm$1.2 & 17.5$\pm$0.9 \\
NGC 7552        & SBab            & 0.79        &  1             & 63--133; 58--138; 68--128    & 19.3$\pm$1.7 & 19.9$\pm$1.1 \\
NGC 7723        & SBb             & 0.66        &  40            & 72--104; 67--109; 77--99     & 16.1$\pm$2.5 & 15.3$\pm$2.0  \\
NGC 7741        & SBcd            & 0.68        &  170           & 165--247; 160--252; 170--242 & 28.5$\pm$2.6 & 30.1$\pm$2.1 \\
\enddata
\tablecomments{Column 1 lists the
galaxy names; Column 2 lists the optical Hubble type from de Vaucouleurs et al.\ 
(1991; hereafter RC3); Column 3 lists the ratio of the minor- to major-axis, 
$b/a$ from NED; Column 4 lists the position angle from RC3; 
Column 5 lists the three radial ranges used for the FFT analysis;
Column 6 lists the
$B$ band pitch angle, $P_B$ and column 7 lists the $H$ band pitch angle,
$P_H$.}
\end{deluxetable*}

Spiral arm pitch angles are measured using the same technique employed
by Seigar et al.\ (2005). A  two-dimensional fast-Fourier decomposition 
technique is used, which employs a program described by Schr\"oder et
al. (1994). Logarithmic spirals are assumed in the
decomposition. The resulting pitch angles are listed in Tables 1,
2 and 3.

\begin{deluxetable*}{llcccll}
\tablecolumns{7}
\tablewidth{7in}
\tablecaption{$B$ band and $K_s$ band spiral arm pitch angles for 9 galaxies from 
our survey.}
\label{tab1a}
\tablehead{
\colhead{Galaxy}     & \colhead{Hubble}          & \colhead{b/a}        & \colhead{PA}       & \colhead{Radial Ranges}             & \colhead{$P_B$}          & \colhead{$P_K$}          \\
\colhead{Name}       & \colhead{Type}            & \colhead{}           & \colhead{}         & \colhead{}                          & \colhead{}               & \colhead{}\\
\colhead{}                & \colhead{}                     & \colhead{}           & \colhead{(degrees)}& \colhead{(arcsec)}                  & \colhead{(degrees)}      & \colhead{(degrees)}      \\
}
\startdata
IC 2522         & SAcd            & 0.71        & 0              & 25--77; 20--82; 30--72    & 38.8$\pm$1.6 & 46.3$\pm$1.2 \\
NGC 1964        & SABb            & 0.38        & 32             & 21--175; 16--180; 26--170 & 13.8$\pm$0.2 & 13.0$\pm$0.7 \\
NGC 2082        & SABc            & 0.94        & 60             & 50--130; 45--135; 55-125  & 27.6$\pm$0.5 & 20.9$\pm$0.6 \\
NGC 2280        & SAcd            & 0.49        & 163            & 55--117; 50--122; 60--112 & 24.2$\pm$1.7 & 22.4$\pm$2.0 \\
NGC 2417        & SABbc           & 0.68        & 81             & 25--71; 20--76; 30--66    & 24.0$\pm$0.7 & 22.2$\pm$3.7 \\
NGC 2935        & SABb            & 0.78        & 0              & 43--93; 38--98; 48--88    & 14.1$\pm$0.2 & 16.5$\pm$1.1 \\
NGC 3318        & SABb            & 0.54        & 78             & 23--89; 18--94; 28--84    & 36.9$\pm$6.5 & 36.9$\pm$2.2 \\
NGC 3450        & SBb             & 0.88        & 140            & 28--48; 25--51; 31--45    & 9.1$\pm$0.4  & 14.5$\pm$0.3 \\
NGC 4050        & SBab            & 0.68        & 85             & 70--91; 67--94; 73--88    & 8.9$\pm$0.7  & 8.8$\pm$1.6  \\
\enddata
\tablecomments{Column 1 lists the
galaxy names; Column 2 lists the optical Hubble type from RC3; 
Column 3 lists the ratio of the minor- to major-axis, 
$b/a$ from NED; Column 4 lists the position angle from RC3; 
Column 5 lists the three radial ranges used for the FFT
analysis; Column 6 lists the $B$ band pitch angle, $P_B$ 
and column 7 lists the $K_s$ band pitch angle, $P_K$.}
\end{deluxetable*}

The amplitude of each Fourier component is given by
\begin{equation}
\label{fft}
A(m,p)=\frac{\sum_{i=1}^{I}\sum_{j=1}^{J}I_{ij}(\ln{r},\theta)\exp{-[i(m\theta _p\ln{r})]}}{\sum_{i=1}^{I}\sum_{j=1}^{J}I_{ij}(\ln{r},\theta)},
\end{equation}
where $r$ and $\theta$ are polar coordinates, $I(\ln{r},\theta)$ is the 
intensity at position $(\ln{r},\theta)$, $m$ represents the number of arms
or modes, and $p$ is the variable associated with the pitch angle $P$, defined
by $\tan{P}=-(m/p)$. Throughout this work we measure the pitch angle $P$ of the
$m=2$ component. The resulting pitch angle measured using equation \ref{fft} 
is in radians, and this is later converted to degrees for ease of perception.

The range of radii over which the Fourier fits were applied were selected to
exclude the bulge or bar (where there is no information about the arms) and 
to extend out to the outer limits of the arms in our images, in such a way that
the 10 kpc radius fell approximately in the middle of this range. 
The radial extent of the bar was measured manually (see e.g.\ Grosbol, Patsis
\& Pompei 2004), and the inner radial limit
applied to the FFT was chosen to be outside this radius. Physical
distances are calculated using a Hubble constant $H_{0}=75$ km s$^{-1}$ 
Mpc$^{-1}$ and recessional velocities from the NASA Extragalactic Database
(NED). Pitch angles are
then determined from peaks in the Fourier spectra, as this is the most powerful
method to find periodicity in a distribution (Consid\`ere \& Athanassoula 1988;
Garcia-Gomez \& Athanassoula 1993). The radial range over which the Fourier
analysis was performed was chosen by eye and is probably the dominant source
of error in the calculation of pitch angles, as spiral arms are only 
approximately logarithmic and sometimes abrupst changes can be seen in spiral 
arm pitch angles (e.g.\ Seigar \& James 1998b) As a result, three radial ranges
were chosen for each galaxy, and a mean pitch angle and standard error 
calculated for every object.

The images were first projected to face-on. Mean uncertainties of position
angle and inclination as a function of inclination were discussed by 
Consid\`ere \& Athanassoula (1988). For a galaxy with low inclination, there
are clearly greater uncertainties in assigning both a position angle and an\
accurate inclination. These uncertainties are discussed by Block et al.\ (1999)
and Seigar et al.\ (2005), who take a galaxy with low inclination 
($<30^{\circ}$) and one with high inclination ($>60^{\circ}$) and varied the
inclination angle used in the correction to face-on. They found that for the
galaxy with low inclination, the measured pitch angle remained the same. 
However, the measured pitch angle for the galaxy with high inclination varied 
by $\pm 10$\%. Since inclination corrections are likely to be largest for
galaxies with the highest inclinations cases where inclination is $>60^{\circ}$
are taken as the worst case scenario.
For galaxies with inclination $i>60^{\circ}$ we take into 
account this uncertainty. Our deprojection method assumes that spiral galaxy
disks are intrinsically circular and flat in nature.

\subsection{Measurement of shear rates}

31 of the galaxies observed here  have H$\alpha$ rotation curve data
measured by Mathewson et al.\ (1992) and Persic \& Salucci (1995). These 
rotation curves are of good quality with an rms error $<10$ km 
s$^{-1}$, and an error associated with folding the two sides of the galaxy
also $<10$ km s$^{-1}$. These rotation curves have been used to estimate the
shear rates in these galaxies, using the same method used by other authors
(e.g. Block et al.\ 1999; Seigar et al.\ 2005; Seigar 2005).

Using equation \ref{shearrate}, we have calculated the shear rates for these
galaxies, over the same radial ranges for which the Fourier analysis was
performed and pitch angles calculated. We have selected several different
radial ranges, just as in the Fourier analysis, and we present mean shear 
rates and standard errors. The dominant sources of error on the shear rate
are the rms error in the rotation curve and the error associated
with folding the two sides of the galaxy. This is typically $<10$\%. In
order to calculate the shear rate, the mean value of $dV/dR$ measured in
km s$^{-1}$ arcsec$^{-1}$ is calculated by fitting a line of constant
gradient to the outer part  of the rotation curve (i.e. past the radius
of turnover and any bar or bulge that may exist in the galaxy). 
Mean shear rates are then calculated from shear rates measured
over three radial ranges, corresponding to the same radial ranges over which
the Fourier analysis was performed. The resulting shear rates are listed in
Table 3.

The choice of using 10 kpc as the radius at which to measure the shear
and spiral arm pitch angle is somewhat arbitrary. The correlation
between these two quantities (Seigar et al.\ 2005 and Figure 3) only
exists when they are measured at a physical radius, chosen indepently
of the disk scalelength. If we measure
pitch angles and shear at a radius chosen using the disk scalelength,
the correlation no longer exists, and the range of shear is 
narrower (between 0.4 and 0.6).
As a result, we chose to measure shear and
pitch angles at a radius independent of the disk scalelength. We
chose 10 kpc (although it could just as easily be 8 kpc or 12 kpc).

The fact that this correlation only exists for a physical radius, suggests
that other quantities may be important in the determination of shear and
pitch angle (e.g.\ the disk scalelength itself or the stability of the disk
as a function of radius). Indeed, when modeling galaxies we use a 1-dimensional
bulge-disk decomposition and use the derived disk scalelength in our codes.

\section{A comparison of optical and near-infrared pitch angles}

\begin{figure}
\plotone{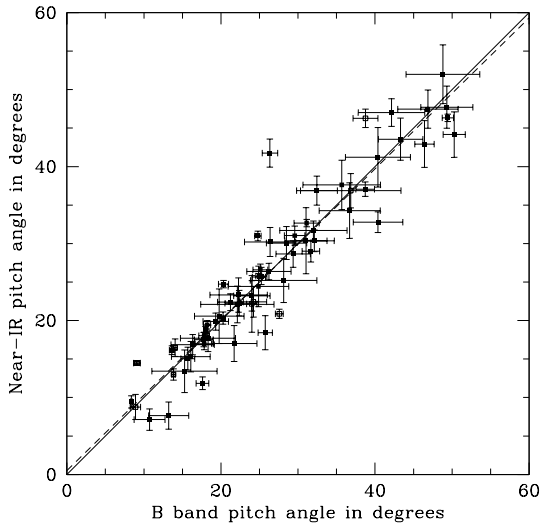}
\caption{A 1:1 correlation between optical spiral arm pitch angle as measured
in the $B$ band and near-infrared spiral arm pitch angle as measured in the
$H$ band for OSUBSGS galaxies and the $K_s$ band for our galaxies. 
This correlation shows that pitch 
angles are similar whether measured in the near-infrared or optical regime.
Only one object shows a difference of $>10^{\circ}$.
The solid squares represent galaxies from the OSUBSGS and the open squares
represent galaxies from our survey. The solid line represents where a 1:1
correlation lies. The dashed line is the best-fit line to the 
data.}
\label{opt-ir-pitch}
\end{figure}

Figure \ref{opt-ir-pitch} shows of plot of $B$ band pitch angle versus
near-infrared pitch angle for the 57 face-on and nearly face-on galaxies taken
from the OSUBSGS and 9  face-on and nearly face-on galaxies taken
from our survey. This plot shows a very tight 1:1 correlation between the 
optical and near-infrared spiral arm pitch angles (correlation 
coefficient=0.95; significance$>$99.99\%), 
showing that pitch angles are similar
whether measured in the optical or near-infrared regimes. In this plot, only
one galaxy has a pitch angle that shows a difference of $>10^{\circ}$ when
measured at $B$ compared with its near-infrared measurement. The outlying
galaxy is NGC 613, which appears to have a strong 4-armed component in
the optical, whereas this is a much weaker feature in the near-infrared.

This result is in contrast to the claims
of Block et al.\ (1999) and Block \& Puerari (1999), who argue that there is
usually a large difference between the spiral arm pitch angle when
measured in the optical and the near-infrared. 
Our result is consistent with the result that Hubble types assigned to
galaxies in the optical and the near-infrared correlate well with each 
other (Eskridge et al.\ 2002).

\begin{figure}
\plotone{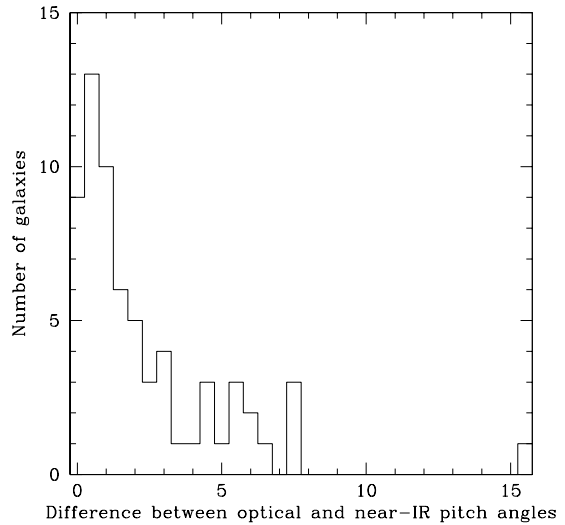}
\caption{Histogram showing the distribution in the absolute difference measured
between the $B$ band and $H$ band pitch angles.}
\label{hist}
\end{figure}

We now analyze the absolute difference 
between the pitch angles measured the the $B$
band and those measured in the near-infrared. The mean of this difference in 
pitch angles is $2\fdg3\pm2\fdg7$, where the error is a standard deviation, 
and this is therefore consistent with the same pitch angle being measured, 
regardless of the waveband. The mode
of the distribution is in the range 0.5$-$1.0 and the median is 1.46.
Furthermore a linear fit to the data in Fig \ref{opt-ir-pitch} yields the
following relationship,
\begin{equation}
P_{NIR}=(0.50\pm1.25)+(0.98\pm0.04)P_{Opt},
\end{equation}
where $P_{NIR}$ is the pitch angle measured in the near-infrared and
$P_{Opt}$ is the pitch angle measured in the optical.
The point at which this line of best fit intersects the $P_{NIR}$-axis is
consistent with zero, and the gradient of the line is consistent with unity,
and so the line of best fit is consistent with a 1:1 correlation between 
spiral arm pitch angles in the optical and near-infrared. It should be noted
that although the difference between optical and near-infrared spiral arm 
pitch angles is consistent with zero from a statistical analysis, minor 
systematic differences cannot be excluded. Small differences between optical
and near-infrared pitch angles have been measured of $1^{\circ}-2^{\circ}$
(e.g.\ Grosbol \& Patsis 1998). Furthermore, color gradients across arms
may be caused by such small differences in pitch angles (e.g. Gonzalez \&
Graham 1996).

Of course, it is still a well-known fact that spiral structure measured
in the near-infrared appears much smoother than that measured in the 
optical (e.g. Block \& Wainscoat 1991; Block et al.\ 1994; Thornley 1996;
Seigar et al.\ 2003). Even in galaxies that appear grand-design in the optical,
small spokes and bifurcations are clearly seen along the length of the dominant
arms. These spokes and bifurcations all but disappear in the near-infrared.
However, our result suggests that the pitch angle of the large-scale spiral
structure remains constant with wavelength in the range 
0.4$\mu$m$<\lambda <$2.2$\mu$m for spirals with a relatively strong
two-armed pattern in the optical. For weaker, more flocculent galaxies there
may well be significant differences due to dust.
This has serious implications for the physical
meaning of the dust-penetrated classification scheme (Block \& Puerari 1999;
Block et al.\ 1999). These implications are discussed in section \ref{next}.

\section{The relationship between shear rate and pitch angle}
\label{next}

\begin{figure}
\plotone{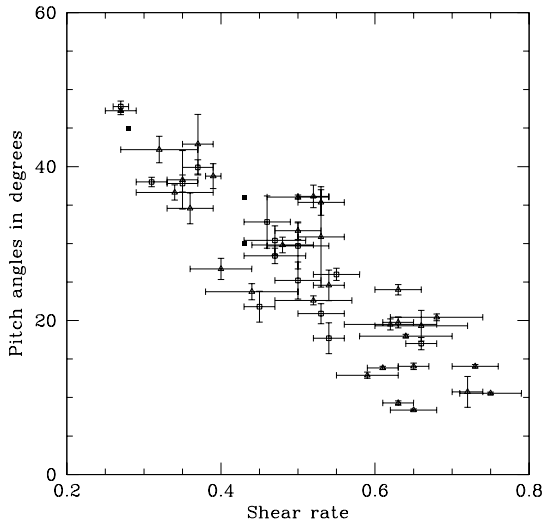}
\caption{Spiral arm pitch angle versus rotation curve shear rate, showing
a strong correlation. The solid squares represent galaxies with data
measured by Block et al.\ (1999), the open squares are galaxies from
Seigar et al.\ (2005), and the open triangles represent the data from
the present sample.}
\label{shear-pitch}
\end{figure}

The correlation between spiral arm pitch angle and rotation curve shear 
rate presented by Seigar et al.\ (2005) only had 17 data points,
including 3 from an earlier study by Block et al.\ (1999). Here we include
a further 31 galaxies with measure pitch angles and shear rates. 
Figure \ref{shear-pitch} shows a plot of shear rate
versus spiral arm pitch angle. A very good correlation still exists
(correlation coefficient=0.89; significance=99.75\%).

The shape of a rotation curve, beyond the turnover radius, is determined
by the amount and distribution of matter contained in a galaxy. The
increase of shear rates from low to high dictates a change in mass 
distributions from small to large central mass concentrations. The correlation
with pitch angle is therefore interpreted as follows: galaxies with
higher rates of shear present a larger central mass concentration and more
tightly wound arms. In contrast, open arm morphologies are associated with
lower rates of galactic shear and lower central mass concentrations. This
correlation was alluded to by the pioneering work of Lin \& Shu (1964) and by
the later spectroscopic study of Burstein \& Rubin (1985). It is in 
agreement with modal theories of spiral structure (e.g. Bertin et al.\ 1989a,
b; Bertin \& Lin 1996) and other numerical models based on the modal 
theory (e.g. Fuchs 1991, 2000).

\begin{deluxetable*}{llcccll}
\tablecolumns{7}
\tablewidth{7in}
\tablecaption{Pitch angles and shear rates for 31 galaxies from our survey.}
\label{tab2}
\tablehead{
\colhead{Galaxy}      & \colhead{Hubble} & \colhead{b/a}  & \colhead{PA}        & \colhead{Radial Ranges}                & \colhead{Pitch}         & \colhead{Shear}         \\
\colhead{Name}        & \colhead{Type}   & \colhead{}     & \colhead{}          & \colhead{}                             & \colhead{Angle}         & \colhead{Rate}\\
\colhead{}                 & \colhead{}            & \colhead{}     & \colhead{(degrees)} & \colhead{(arcsec)}                     & \colhead{(degrees)}           & \colhead{}              \\
}
\startdata
ESO 9-G10   & SAc     & 0.76 & 171	 & 27--101; 22-106; 32--96      & 23.7$\pm$1.1 & 0.44$\pm$0.06 \\		
ESO 121-G26 & SBc     & 0.63 & 115	 & 51--85; 46--90; 56--80       & 10.5$\pm$0.2 & 0.75$\pm$0.04 \\
ESO 582-G12 & SAc     & 0.64 &  48	 & 32--102; 27--107; 37--97     & 22.6$\pm$0.6 & 0.52$\pm$0.05 \\
IC 2522     & SAcd    & 0.71 &   0	 & 25--77; 20--82; 30--72       & 38.8$\pm$1.6 & 0.39$\pm$0.01 \\		
IC 2537     & SABc    & 0.65 &  26	 & 30--82; 25--87; 35--77       & 9.3$\pm$0.3  & 0.63$\pm$0.02 \\
IC 3253     & SAc     & 0.39 &  23	 & 22--92; 17--97; 27--87       & 24.6$\pm$2.0 & 0.36$\pm$0.03 \\
IC 4538     & SABc    & 0.77 &  50	 & 42--66; 37--71; 47--61       & 38.3$\pm$3.8 & 0.35$\pm$0.02 \\
IC 4808     & SAc     & 0.42 &  45	 & 15--45; 10--50; 20--40       & 14.1$\pm$0.4 & 0.65$\pm$0.02 \\
NGC 150     & SBbc    & 0.50 & 118	 & 70--126; 65--131; 75--121    & 8.4$\pm$0.1  & 0.65$\pm$0.03 \\
NGC 151     & SBbc    & 0.42 &  75	 & 30--52; 27--55; 33--49       & 36.1$\pm$1.5 & 0.52$\pm$0.02 \\
NGC 578     & SABc    & 0.63 & 110	 & 65--125; 60--130; 70--120    & 18.0$\pm$0.2 & 0.64$\pm$0.06 \\
NGC 908     & SAc     & 0.43 &  75	 & 56--150; 51--155; 61--145    & 12.9$\pm$0.4 & 0.59$\pm$0.04 \\
NGC 1232    & SABc    & 0.88 & 108	 & 53--141; 48--146; 58--136    & 19.3$\pm$2.0 & 0.66$\pm$0.06 \\
NGC 1292    & SAc     & 0.43 &   7	 & 53--173; 48--178; 58--168    & 29.8$\pm$1.0 & 0.48$\pm$0.04 \\
NGC 1300    & SBbc    & 0.66 & 106	 & 80--116; 75--121; 85--111    & 31.7$\pm$1.1 & 0.50$\pm$0.03 \\
NGC 1353    & SAbc    & 0.41 & 138	 & 63--139; 58--144; 68--134    & 36.6$\pm$1.0 & 0.34$\pm$0.05 \\
NGC 1365    & SBb     & 0.55 &  32	 & 61--129; 56--134; 66--124    & 35.4$\pm$1.7 & 0.53$\pm$0.03 \\
NGC 1559    & SBcd    & 0.57 &  64	 & 82--156; 77--161; 87--151    & 20.4$\pm$0.4 & 0.68$\pm$0.06 \\
NGC 1566    & SABbc   & 0.80 &  60	 & 73--133; 68--138; 78--128    & 36.0$\pm$0.3 & 0.50$\pm$0.04 \\
NGC 1964    & SABb    & 0.38 &  32	 & 21--175; 16--180; 26--170    & 13.8$\pm$0.2 & 0.61$\pm$0.02 \\
NGC 2280    & SAcd    & 0.50 & 163	 & 55--117; 50--122; 60--112    & 24.2$\pm$1.7 & 0.32$\pm$0.05 \\
NGC 2417    & SABbc   & 0.68 &  81	 & 25--71; 20--76; 30--66       & 24.0$\pm$0.7 & 0.63$\pm$0.03 \\
NGC 2835    & SABc    & 0.67 &   8	 & 110--240; 105--245; 115--235 & 19.5$\pm$0.7 & 0.62$\pm$0.06 \\
NGC 2935    & SABb    & 0.78 &   0	 & 43--93; 38--98; 48--88       & 14.1$\pm$0.2 & 0.73$\pm$0.03 \\
NGC 3052    & SABc    & 0.65 & 102	 & 25--57; 20--62; 30--52       & 19.8$\pm$0.7 & 0.63$\pm$0.02 \\
NGC 3054    & SABbc   & 0.61 & 118	 & 44--84; 39--89; 49--79       & 42.9$\pm$3.9 & 0.37$\pm$0.02 \\
NGC 3223    & SAbc    & 0.61 & 135	 & 32--74; 27--79; 37--69       & 10.7$\pm$2.0 & 0.72$\pm$0.02 \\
NGC 3318    & SABb    & 0.54 &  78	 & 23--89; 18--94; 28--84       & 36.9$\pm$6.5 & 0.53$\pm$0.03 \\
NGC 5967    & SABc    & 0.59 &  90	 & 46--76; 41--81; 51--71       & 47.3$\pm$0.5 & 0.27$\pm$0.02 \\
NGC 7083    & SABc    & 0.59 &   5	 & 32--78; 29--81; 35--75       & 26.7$\pm$1.4 & 0.40$\pm$0.04 \\
NGC 7392    & SBab    & 0.62 & 123	 & 30--66; 25--71; 35--61       & 24.6$\pm$2.0 & 0.54$\pm$0.02 \\
\enddata
\tablecomments{Column 1 lists the galaxy 
names; Column 2 lists their Hubble types from RC3; Column 3 lists the 
ratio of the minor- to major-axis from NED; Column 4 lists the position 
angle from RC3; Column 5 lists the three radial ranges used for the 
FFT analysis; Column 6 lists the pitch angle and Column 7 lists the 
shear rate.}
\end{deluxetable*}

The correlations shown in Figures \ref{opt-ir-pitch} and \ref{shear-pitch}
have significant implications for the dust-penetrated classification 
scheme introduced by Block \& Puerari (1999). These authors argue that
their dust-penetrated classification scheme is needed, because spiral
arm pitch angles are different in the near-infrared when compared to the
optical, and as a result they classify galaxies into three bins, based
on their pitch angles when measured in the near-infrared. 
However, we have shown (in Figure \ref{opt-ir-pitch}) that spiral galaxies with
a relatively strong two-armed spiral pattern, have optical and near-infrared 
pitch angles that are nearly always very similar.
This, therefore, seems to lessen the need for the dust-penetrated
classification scheme. In Figure \ref{shear-pitch}, however, we show that
a correlation still exists between shear rate and spiral arm pitch angle,
no matter what waveband the pitch angle is measured in. This suggests that
classifying galaxies based on their pitch angles has some physical basis,
because it tells us something about their central mass concentrations. 

Given the correlation between shear rate and pitch angle we derive the 
following expression relating these terms
\begin{equation}
\label{shearpitch}
P=(64.25\pm2.87)-(73.24\pm5.53)S,
\end{equation}
where $P$ is the pitch angle in degrees and $S$ is the shear rate.

\section{Connecting pitch angles to central mass concentrations}

Our goal now is to demonstrate how spiral arm pitch angle measurements
coupled with a  bulge-disk decomposition can  be used to constrain the
overall mass distribution in  spiral galaxies.  We outline the  method
for two example galaxies with pseudo-bulges (see Kormendy \& Kennicutt
2004 for a review about  pseudo-bulges). 
The two galaxies we chose are chosen such that one galaxy, ESO 582-G12,
has a flat rotation curve, with a shear, $S=0.52\pm0.05$ in the middle 
of the range, and the other galaxy, IC 2522, has a continually rising
rotation curve, with a shear, $S=0.39\pm0.01$, towards the low end
of the range of shears.
As a normalization for each galaxy we take its measured
rotation speed at 2.2 disk scale lengths, $V_{2.2}$ (listed in the last
column of Table 4), and include a bulge-disk decomposition
in order to estimate the baryonic contribution to each rotation curve.
In  future work paper we plan to use the same
methodology for a large sample of galaxies.

\begin{figure*}
\vspace*{3in}
%\plottwo{seigar_f4a.eps}{seigar_f4b.eps}
\plottwo{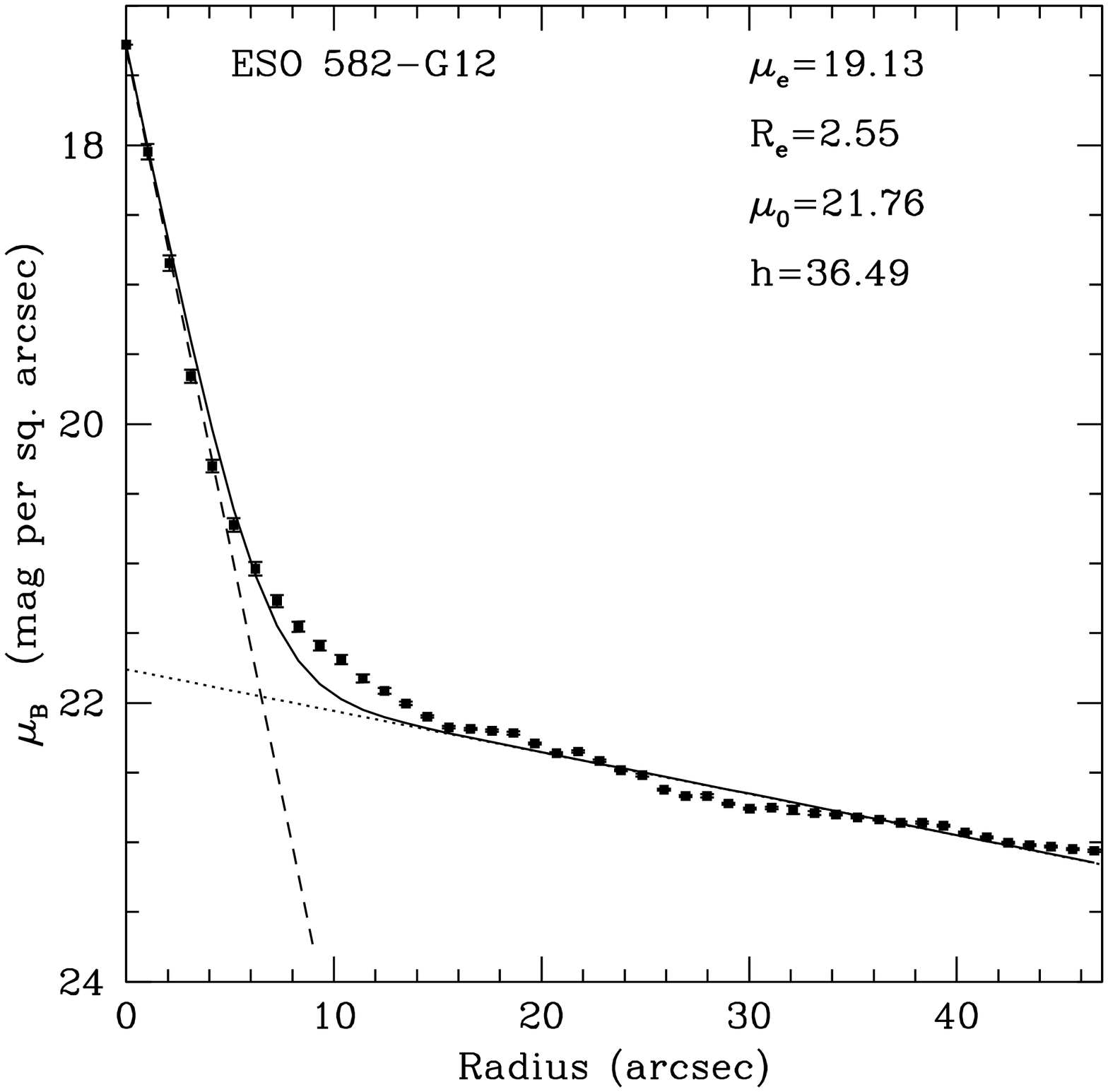}{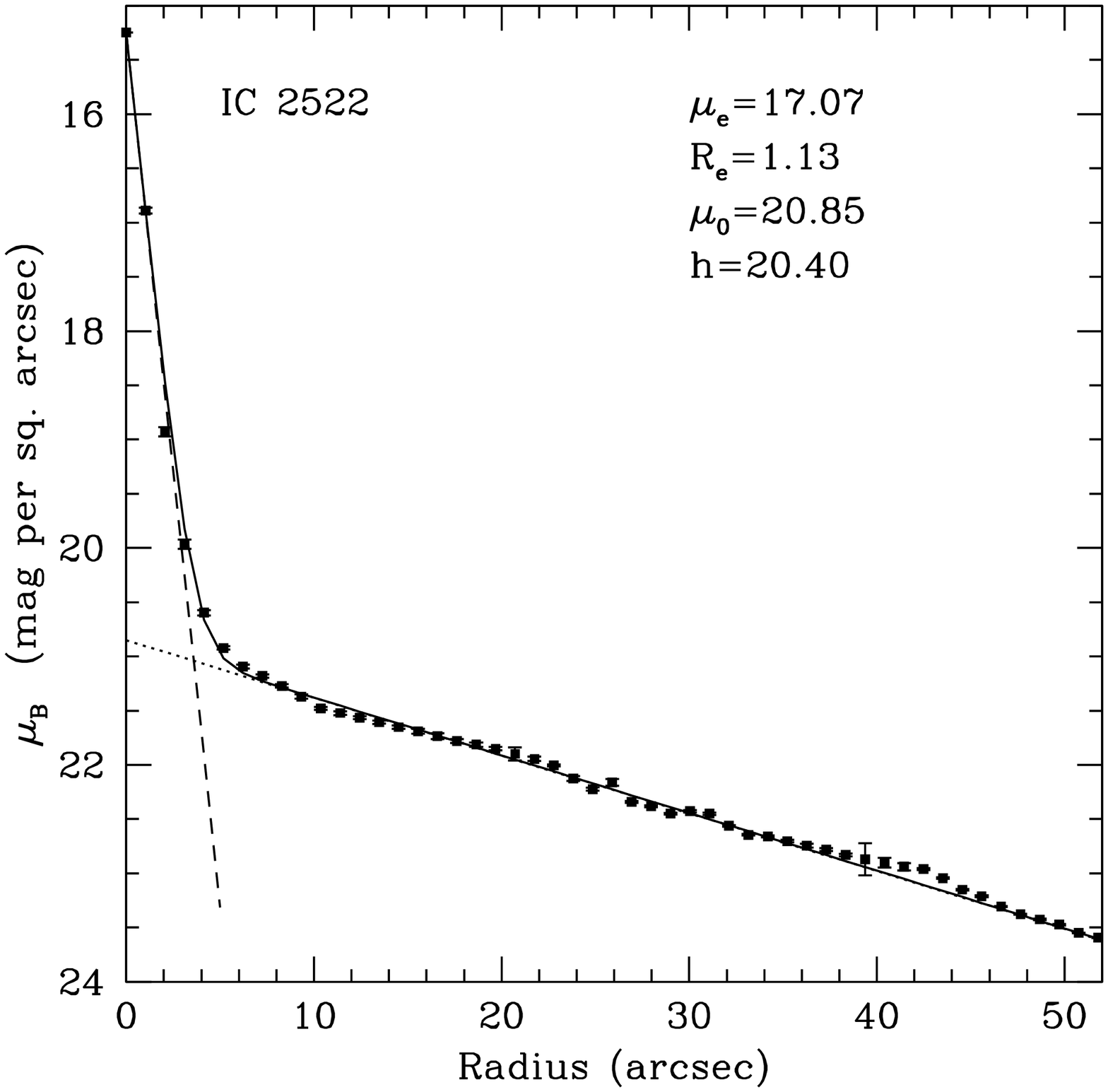}
\caption{Bulge-disk decomposition for two example spiral galaxies
as observed in the B-band. The top
row shows a B band image of each galaxy and the bottom row
shows the surface brightness profile and results of the 
decomposition. The effective radii $R_e$ and disk 
scalelengths $h$ are measured in arcsec. The dashed line is the
best fit bulge model, the dotted line is the best fit disk model
and the solid line is the best combined bulge+disk model.
{\em Left panel}: ESO 582-G12. {\em Right Panel}: IC 2522.
The images are aligned such that North is up and the field is
4.5$\times$4.5 arcmin.}
\label{decompose}
\end{figure*}

For each of our galaxies we produce surface brightness profiles
using the IRAF {\bf Ellipse} routine, which fits ellipses to
isophotes in an image, using an iterative method described
by Jedrzejewski (1987). From the surface brightness profile
we then determine the disk and bulge B-band luminosity
using  an exponential disk and an exponential (pseudo) bulge. In future
papers a more general S\'ersic parameterization will be used for the bulge
component. We utilize 
a 1-D bulge-disk decompostion routine, which performs Levenberg-Marquardt
least-squares minimization.
Explicity we fit an exponential profile for each bulge
via 
\begin{equation}
\mu(R)=\mu_e \exp{(-1.679[(R/R_e)-1])},
\label{bulge}
\end{equation}
where $R_e$ is the effective radius, containing 50\% of the total light of the
bulge and $\mu_e$ is the surface brightness at $R_e$.
We fit the disk component using
\begin{equation}
\mu(R)=\mu_0 \exp{(-R/h)},
\label{disk}
\end{equation}
where $\mu_0$ is the central surface brightness and $h$ 
is the scalelength of the disk.
The results of this bulge-disk decompostion can be seen in Figure 
\ref{decompose} and Table 4.

We assign masses  to the disk  and bulge components  using a  range of
stellar mass-to-light ratios from Bell \& de Jong
(2001). Specifically, in our rotation curve models we allow
mass-to-light  ratios of $(M/L) =   1.0,    1.3$  and  $1.6$ 
(measured in B band solar units), and use our
photometrically-derived disk  and bulge  light  profiles 
$L_{B} = L_{disk} + L_{bulge}$
 to determine the stellar mass
contributuion to each rotation curve: 
$M_{*} = (M/L) L_{B}$.

We now explore a range of allowed dark matter halo
masses and density profiles, adopting two 
 extreme models for disk galaxy formation.
In the first, we assume that the dark matter halos
surrounding these galaxies have not responded significantly
to the formation of the disks, i.e. adiabatic contraction (AC)
does not occur.  We refer to this as
our ``non-AC'' model.
 In this case, the dark matter
contributuion to each galaxy rotation curve 
is described by a density profiles that mirrors those found
in {\rm dissipationless} dark matter simulations:
\begin{equation}
\rho(r) = \frac{\rho_s}{(r/r_s)\left(1+r/r_s\right)^2},
\label{eqt:nfw}
\end{equation}
where $r_s$ is  a    characteristic ``inner" radius, and     $\rho_s$ a
corresponding  inner density.  Here we have adopted the profile
shape of Navarro, Frenk \& White (1996; hereafter NFW).  
The NFW profile is a two-parameter function
and is completely specified by choosing two independent paramters,
e.g. the virial mass $M_{\rm vir}$ (or virial radius $R_{\rm vir}$)
and concentration $c_{\rm vir} \equiv R_{\rm vir}/r_s$ define
the profile completely (see Bullock et al.\ 2001b,
for a discussion).  Similarly, given a virial mass
$M_{\rm vir}$ and the dark matter circular velocity at any radius, the
halo concentration $c_{\rm vir}$ is completely determined.

%JB modified

In the second  class of  models we adopt  the scenario  of ``adiabatic
contraction''  (AC) discussed  by Blumenthal  et  al.  (1986; see also
Bullock et  al. 2001a, Pizagno et al.\  2005).  Here we assume that the
baryons and dark matter initially follow an NFW  profile and that the
baryons cool  and  settle into the   halo center slowly compared  to a
typical  orbital  time.   This   slow  infall provokes   an  adiabatic
contraction  in  the  halo  density distributuion and   
gives rise to a  more  concentrated  dark  matter profile.  
The idea of adiabatic contraction was originally discussed
as to explain the ``conspiracy'' between dark halos and
disk sizes which gives rise to a featureless rotation curve 
(Rubin et al.\ 1985) but has since proven to be remarkably
accurate in describing the formation of disk galaxies in 
numerical simulations (e.g. Gnedin et al.\ 2004, and references therein),
although the degree to which this process operates in the real universe
is currently uncertain.  For example, Dutton et al.\ (2005), 
showed that adiabatic contraction models are inconsistent with the
rotation curves measured, and the expected NFW concentrations, for a 
sample of 6 galaxies. They suggest that mechanisms such as stellar
feedback and stellar bars may result in less concentrated halos than
predicted by adiabatic concentration.

In our AC model, we take the
contraction into  account following the  prescription of Blumenthal et
al. (1986).  Note that  Gnedin   et al.\   (2004) advocate  a  slightly
modified prescription, but the differences between the two methods are
small compared to  the differences between  our  ``AC'' model and  our
non-AC   model.    In principle, any    observational  probe  that can
distinguish  between AC and non-AC type scenarios
provides an important constraint on 
the nature of gas infall into galaxies (i.e.  was it fast or
was it slow?).

For each galaxy we iterate over the  central and $\pm 1-\sigma$ values
found  in the bulge-disk decompositions  for  $h$ and $L_{disk}$ and
explore the three values of mass-to-light ratio discussed above $(M/L)
=   1.0,  1.3,$ and  $1.6$.  In each case we assume that $M/L$ remains
constant with radius. For  each  choice  of bulge-disk  model
parameters and mass-to-light ratios  we  allow the (initial) halo  NFW
concentration parameter  to  vary  over  the range of   viable values:
$c_{\rm vir} =  3-31$ (Bullock et al.\  2001b).  We then determine  the
halo virial mass $M_{\rm  vir}$ necessary  to  reproduce the  value of
$V_{2.2}$ for the galaxy and determine the implied fraction of the mass
in the system in the form of stars compared to that 
``expected'' from the Universal baryon fraction: 
$f_* \equiv  M_*/(f_b M_{\rm vir})$.  We
make the (rather  loose) demand  that $f_*$ lies  within  the range of
plausible values $f_* = 0.01 - 1 f_b$.

\begin{deluxetable*}{llllllc}
\tablecolumns{7}
\tablewidth{7in}
\tablecaption{Galaxy parameters used in model constraints}
\label{tab4}
\tablehead{
\colhead{Galaxy Name}    & \colhead{Distance}        & \colhead{$h$}               & \colhead{$h$}              & \colhead{B/D}      & \colhead{$L_{disk}$}            & \colhead{$V_{2.2}$}     \\
\colhead{}               & \colhead{(Mpc)}           & \colhead{(arcsec)}          & \colhead{(kpc)}            & \colhead{}         & \colhead{($\times 10^{10}L_{\odot}$)} 	& \colhead{(km s$^{-1}$)}  \\
}
\startdata
ESO 582-G12              & 31.0$\pm$0.1              & 36.5$\pm$3.8              & 5.48$\pm$0.57              & 0.106              & 1.27$\pm$0.11            	& 145	 \\
IC 2522                  & 40.3$\pm$0.1              & 20.4$\pm$2.1              & 3.98$\pm$0.41              & 0.164              & 1.55$\pm$0.14             	& 216	\\
\enddata
\tablecomments{Column 1 lists the galaxy names;
Column 2 lists the distances to the galaxies in Mpc, calculated using a Hubble constant $H_0=75$ km s$^{-1}$ Mpc$^{-1}$;
Column 3 lists the disk scalelengths, $h$, in arcsec;
Column 4 lists the disk scalelengths, $h$, in kpc;
Column 5 lists the bulge-to-disk ratio , B/D;
Column 6 lists the disk B-band luminosity, $L_{disk}$;
and Column 7 lists the rotational velocity at 2.2 disk scalelength, $V_{2.2}$.}
\end{deluxetable*}

\begin{figure*}
\hspace*{0.5cm}
\includegraphics[width=5.4cm]{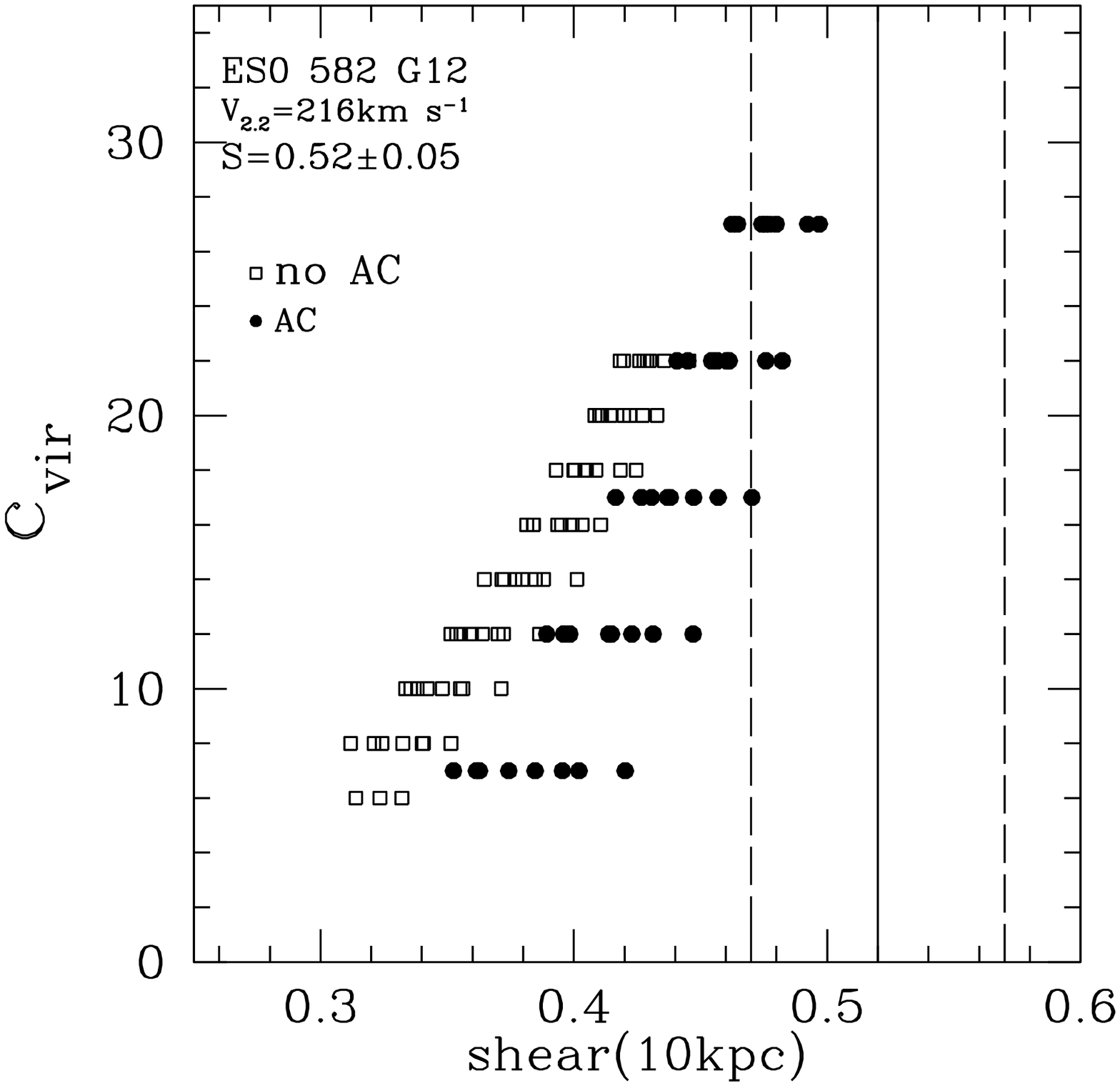}
\includegraphics[width=5.4cm]{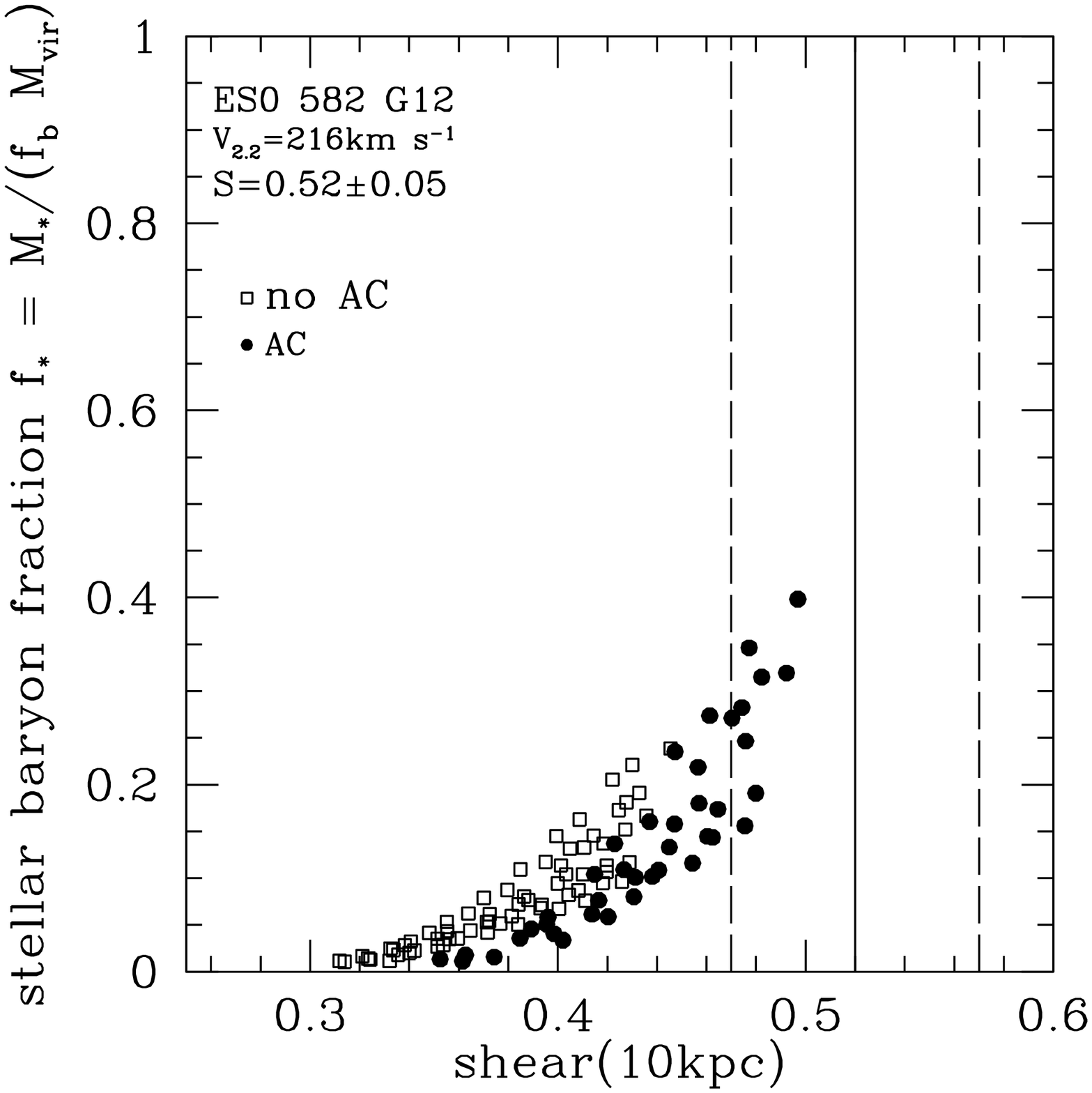}
\includegraphics[width=5.4cm]{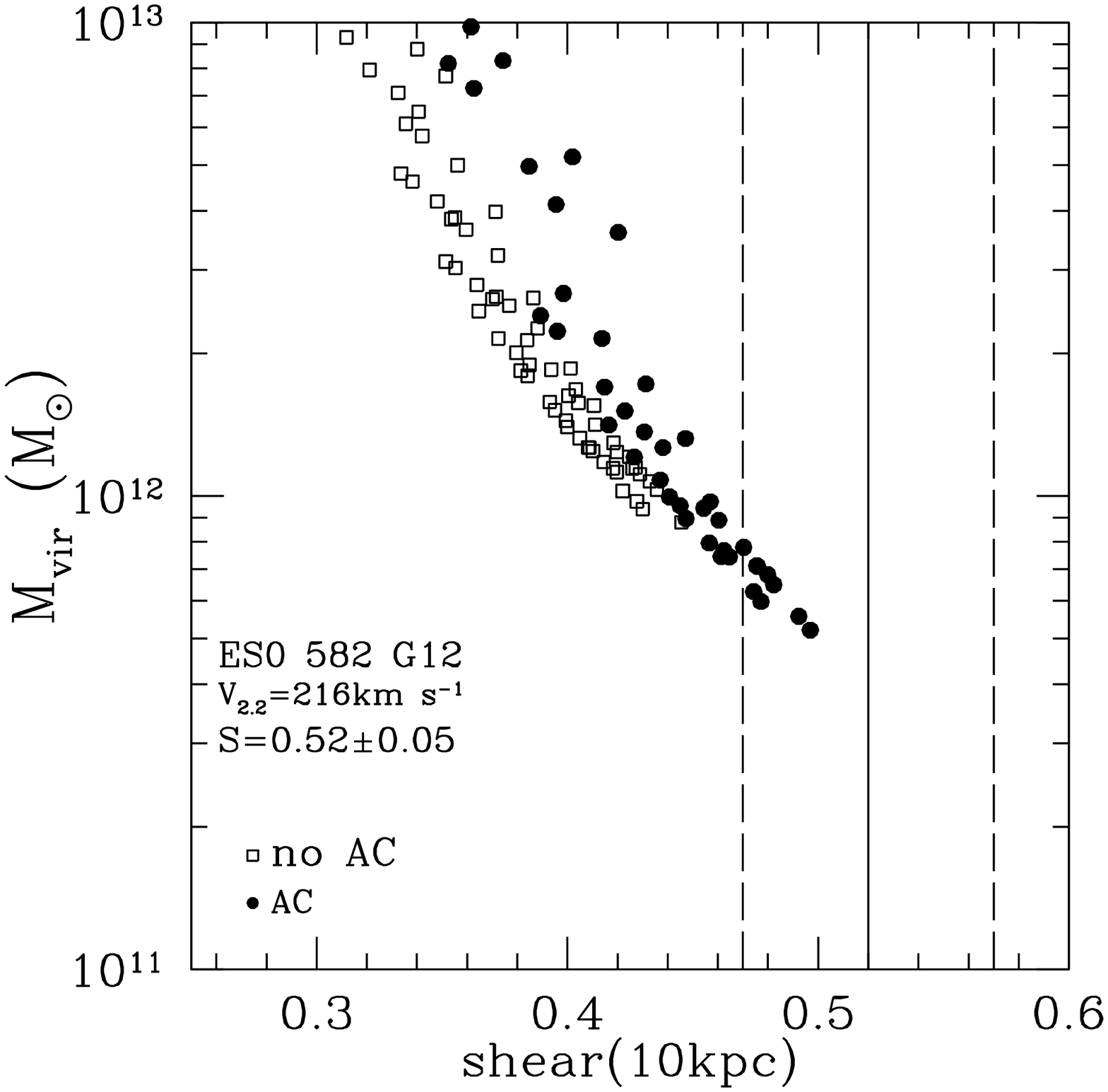}
\caption{Model Results for ESO 582-G12. 
{\em Left}: NFW concentration  versus shear, {\em center}:
stellar baryon fraction  versus shear and {\em right}: virial
mass versus shear. The open squares represent the model without
an adiabatically contracted dark matter halo 
(non-AC) and the solid circles represent the model with adiabatic
contraction (AC). The vertical lines present the measured shear (solid
line) with the 1-$\sigma$ error added and subtracted (dashed lines).}
\label{eso582_m}
\end{figure*}

\begin{figure*}
\hspace*{0.5cm}
\includegraphics[width=5.4cm]{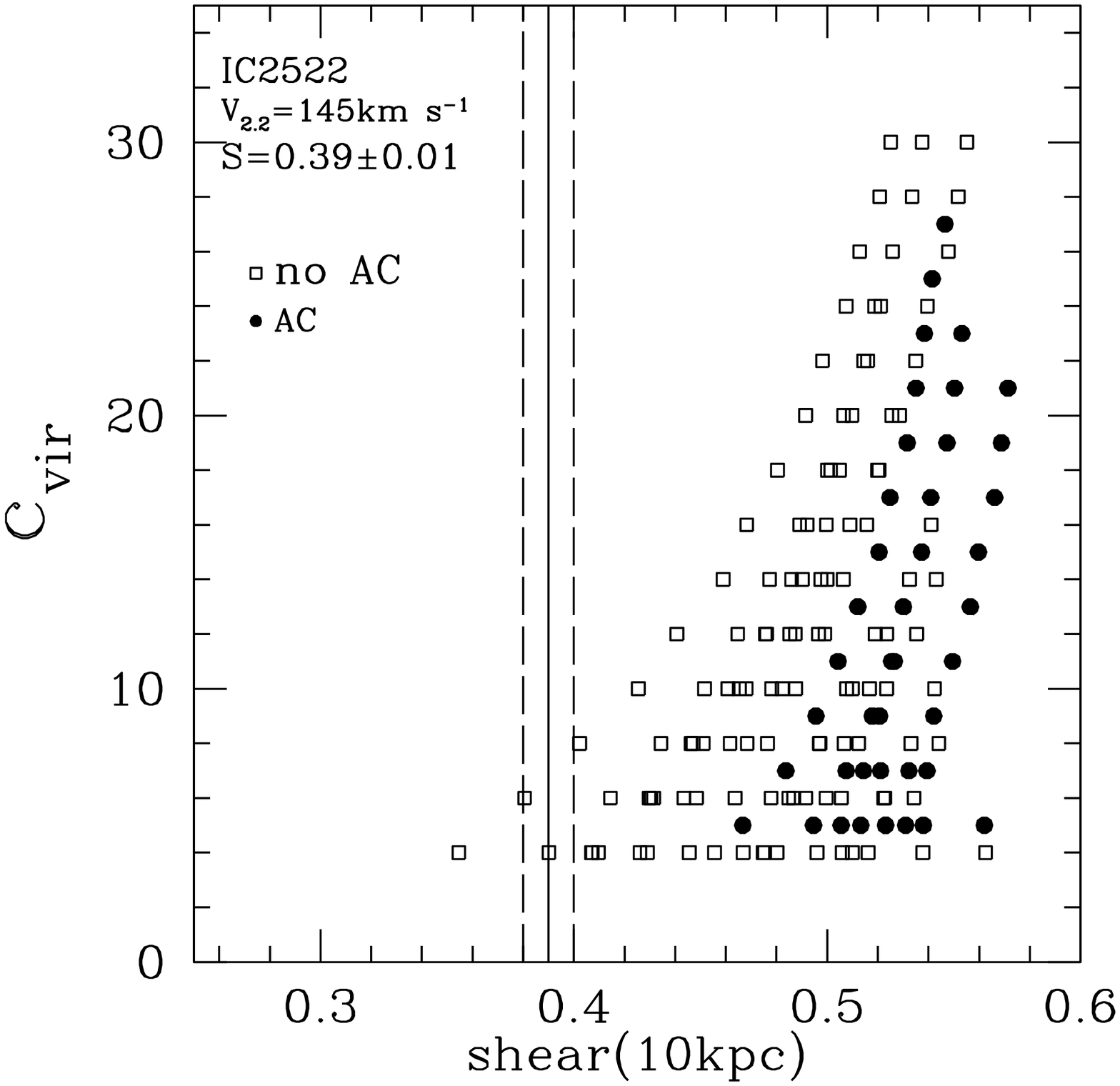}
\includegraphics[width=5.4cm]{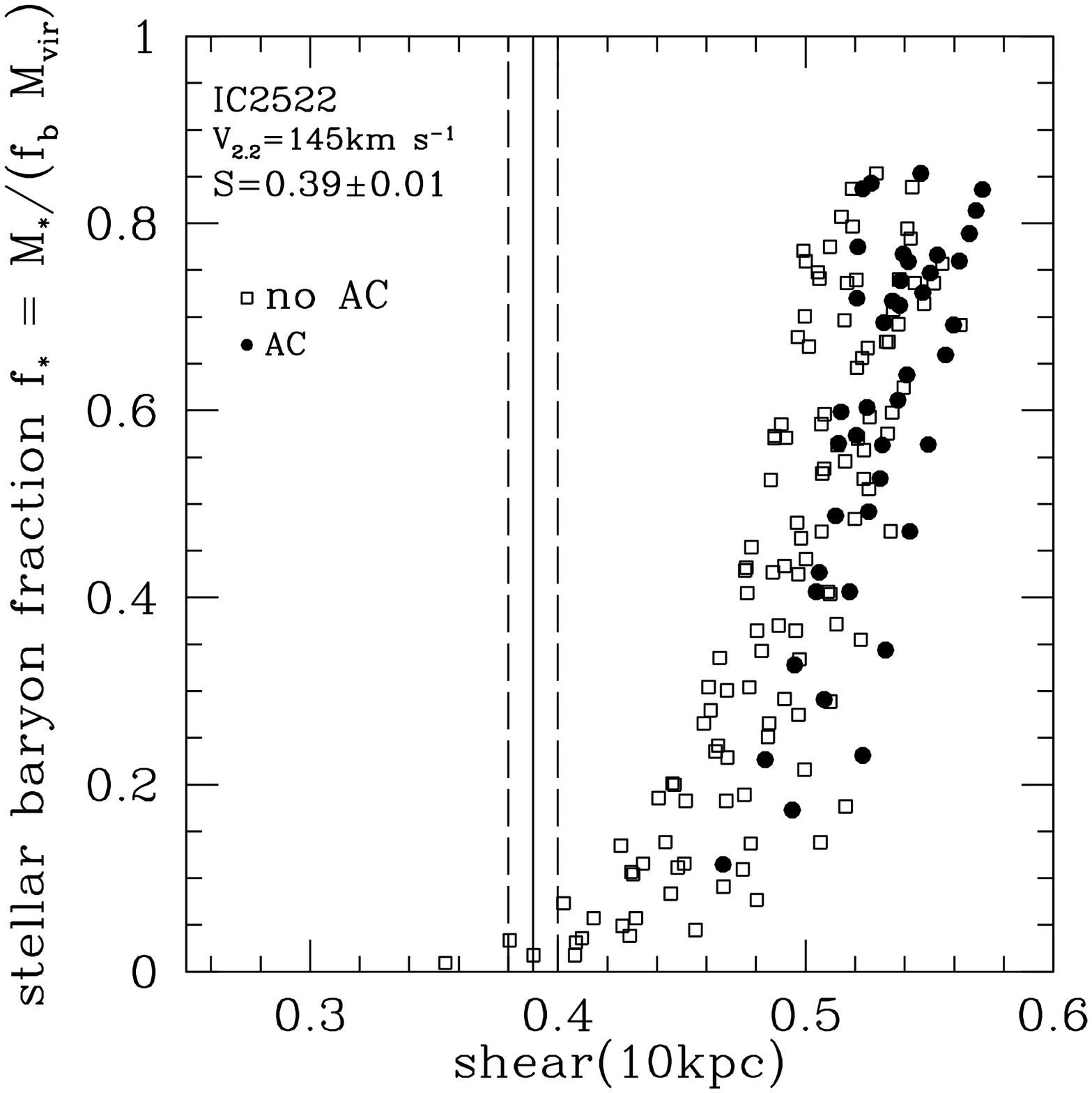}
\includegraphics[width=5.4cm]{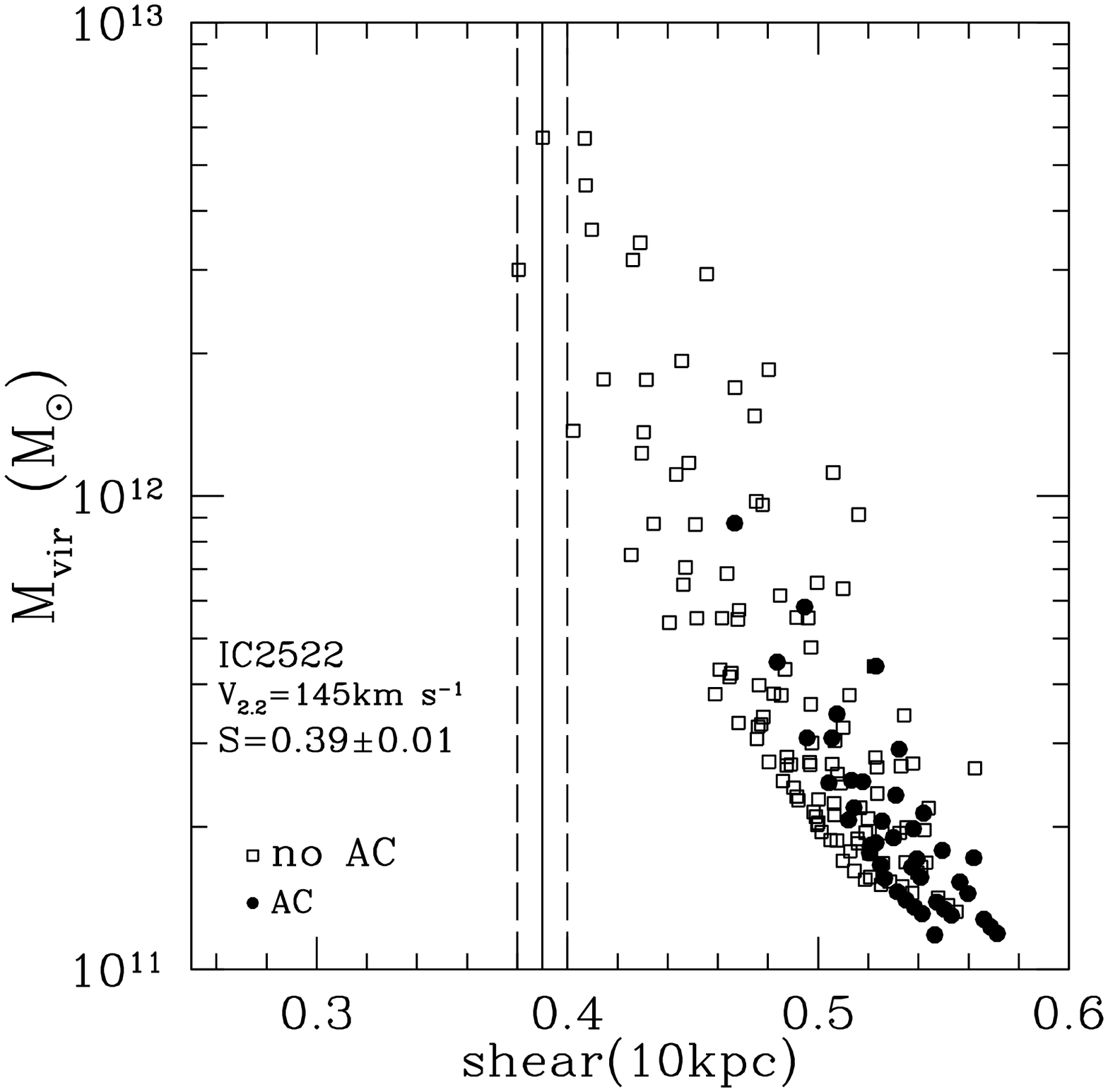}
\caption{Model Results for IC 2522.
{\em Left}: NFW concentration  versus shear, {\em center}:
stellar baryon fraction  versus shear and {\em right}: virial
mass versus shear. The open squares represent the model without
an adiabatically contracted dark matter halo 
(non-AC) and the solid circles represnt the model with adiabatic
contraction (AC). The vertical lines present the measured shear 
(solid line) with the 1-$\sigma$ error added and subtracted
(dashed lines).}
\label{ic2522_m}
\end{figure*}

For each  chosen value of  $c_{\rm  vir}$  and adopted disk  formation
scenario (AC or non-AC)  the   
chosen $V_{2.2}$ constraint defines  the
rotation curve completely, and thus provides  an implied shear rate at
every radius.  The  three  panels    of Figures \ref{eso582_m}     and
\ref{ic2522_m} show the results of this exercise for ESO  582-G12 and IC
2522 respectively.  In each  panel, open symbols   are for the  non-AC
model and the solid symbols reflect the AC assumption.  Each point
represents a distinct input combination of $h$, $L_{disk}$, and
$(M/L)$.  The measured shear rate is illustrated by a solid vertical line
and the $\pm 1-\sigma$ range in the observe shear rate for
each galaxy is shown by the two vertical dashed lines in each panel.

Consider first the left hand panel of Figure \ref{eso582_m}.  Here we
plot the dark matter halo concentration parameter versus the shear
rate measured at $10$ kpc. More concentrated halos generally produce
higher shear rates, as expected.  It can be seen
that for a given NFW concentration, $c_{vir}$, several values of shear
are possible. This is due to changes in the baryon contrbution (i.e.
the disk mass and disk scalelength) to the rotation curve. Whether
an increase in the baryon contribution causes the shear to increase
or decrease depends on the size of the disk (i.e. the disk scalelength).
The same is true for Figure 6.
In the AC model (solid), $c_{\rm vir}$ refers to the halo
concentration {\em before} the halo is adiabatically contracted.
This is why the AC points tend to have higher shear values at fixed
$c_{\rm vir}$ compared to the non-AC case.  Every point
(or $h$, $L_{disk}$, $(M/L)$ input combination) has an associated
stellar baryon fraction $f_*$ and dark halo virial mass $M_{\rm vir}$.
These values are plotted versus shear rate in the middle and right-hand
panels in Figure \ref{eso582_m}.  Observe that with only the
$V_{2.2}$ constraint imposed (all points), then a wide range of
dark matter halo properties are allowed.  Once we constrain the models
by forcing the predicted shear to be consistent the observed
range $S=0.47-0.57$,  we favor models that include adiabatic contraction.
The dark matter of halo of ESO 582-G12 is constrained to be of relatively
low mass, $M_{\rm vir} \simeq 5-8 \times 10^{11}$M$_{\odot}$, and
to have a fairly high NFW concentration, $c_{\rm vir} > \sim  16$.  This is consistent
with the high-end of the the $c_{\rm vir}$ distribution predicted for
LCDM \citep{bullock01b}.  The stellar mass in the galaxy is
between $\sim 15$ and $40$\% of the the baryonic mass associated with
its dark matter halo.

Figure \ref{ic2522_m} shows analogous results for IC 2522.  Unlike ESO
582 G12, this galaxy  prefers a more massive, low-concentration  halo:
$M_{\rm vir} \simeq 1.5-6 \times 10^{12}$M$_{\odot}$ with $c_{\rm vir}
\simeq  2-8$,  at the    low  end of   the $c_{\rm  vir}$ distribution
\citep{bullock01b}.  The implied stellar mass is quite small: $< 10$\%
of the universal baryon budget has ended up on stars. Furthermore, the
measured shear rate of this galaxy favors our non-AC model. This 
is in qualitative 
agreement with results presented by Dutton et al.\ (2005).

%JB: I moved this to the discussion above

%%%% Above I have included reference to Dutton et al. (2005) who favor non-AC (this is the Courteau paper, he's the 2nd author)

\begin{figure*}
\plottwo{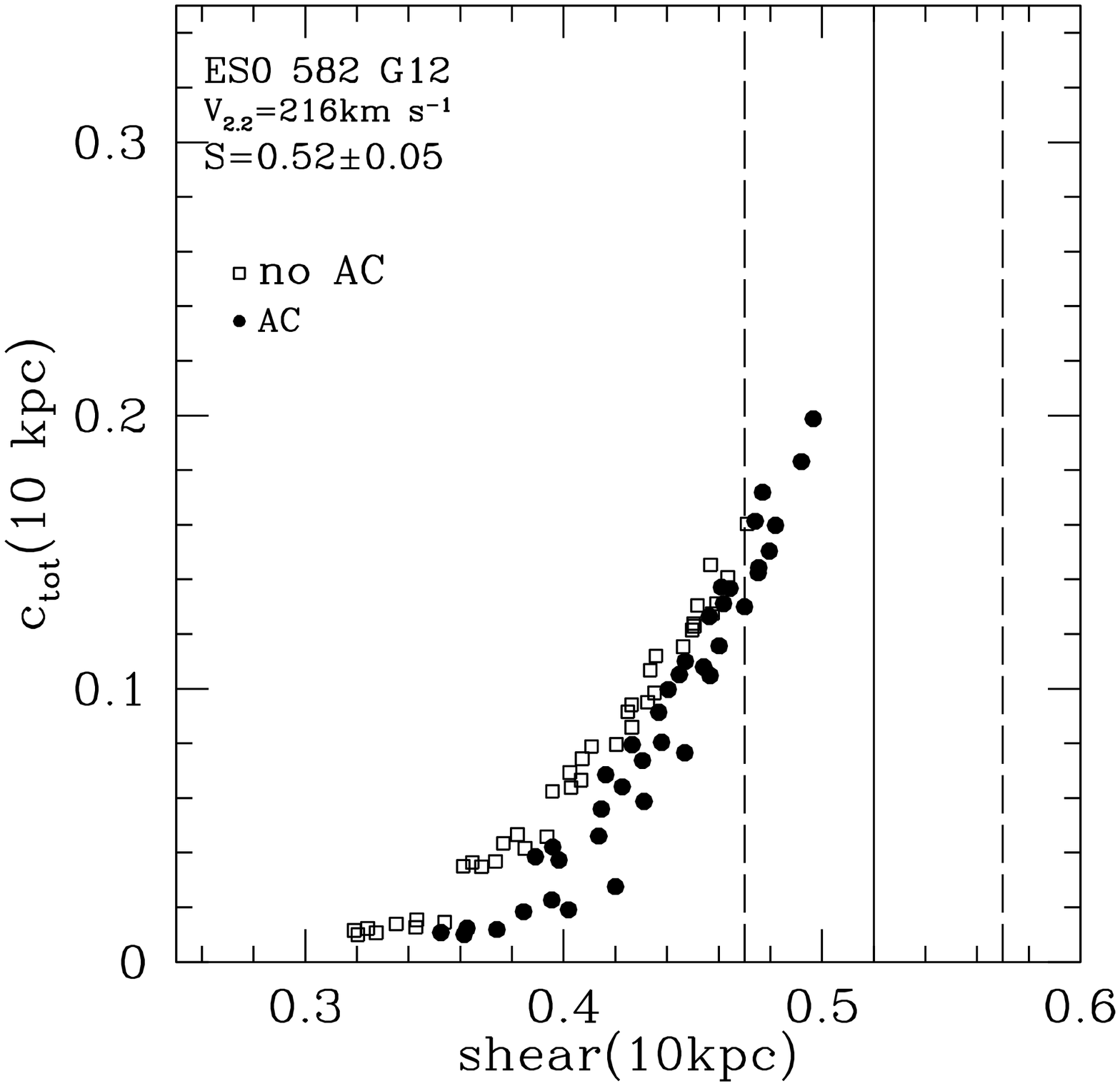}{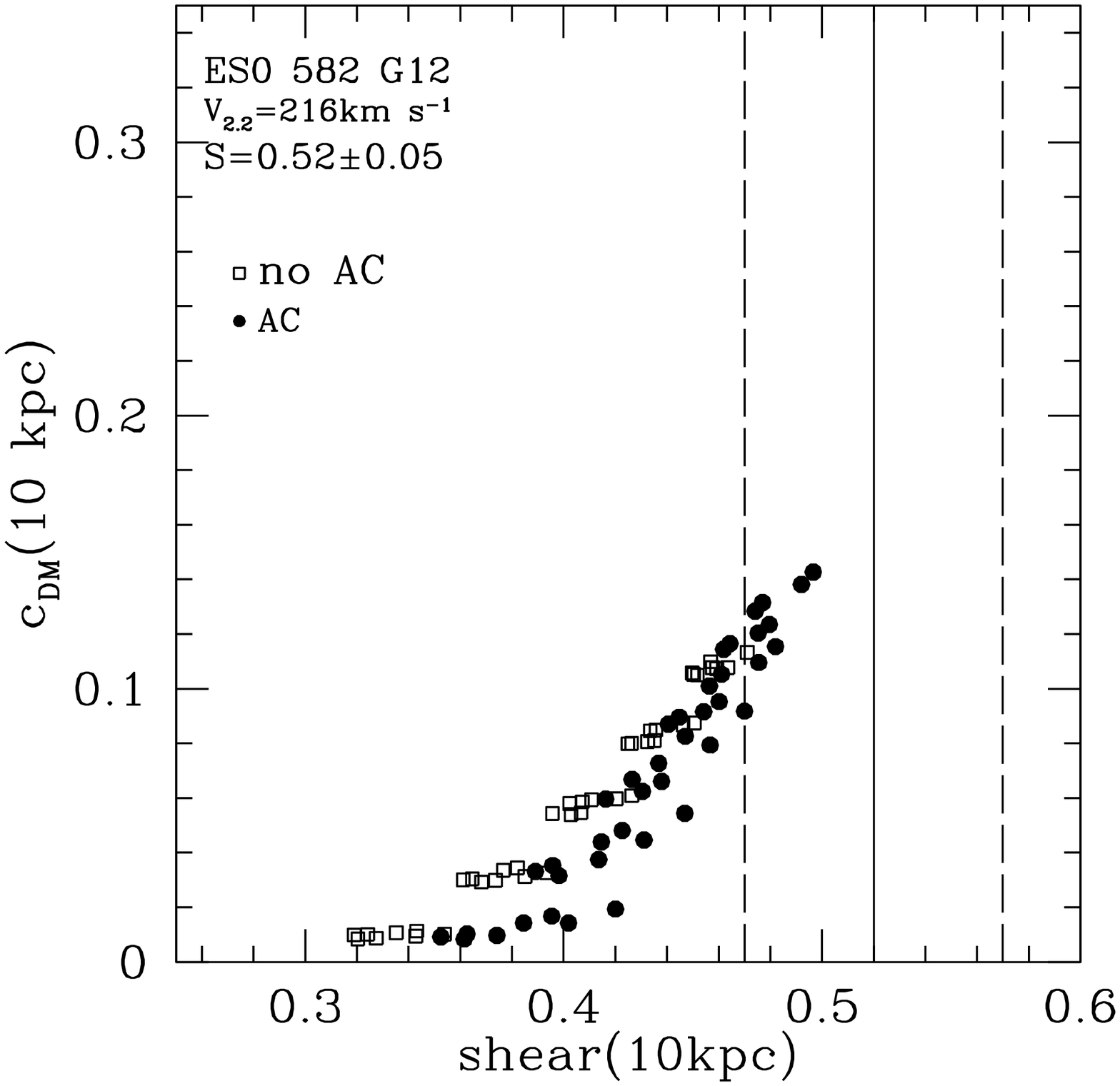}
\caption{Model results for
ESO 582-G12: {\em Left}: Central mass concentration (mass fraction within 
$10$ kpc) versus shear, {\em right}:
dark matter concentration (dark matter mass fraction within $10$ kpc)
versus shear. The open squares represent the model without
adiabatic contraction of the stellar halo
and solid circles represnt the model with adiabatic
contraction. The vertical lines present the measured shear (solid line)
with the 1-$\sigma$ error added and subtracted (dashed lines).}
\label{eso582}
\end{figure*}

\begin{figure*}
\plottwo{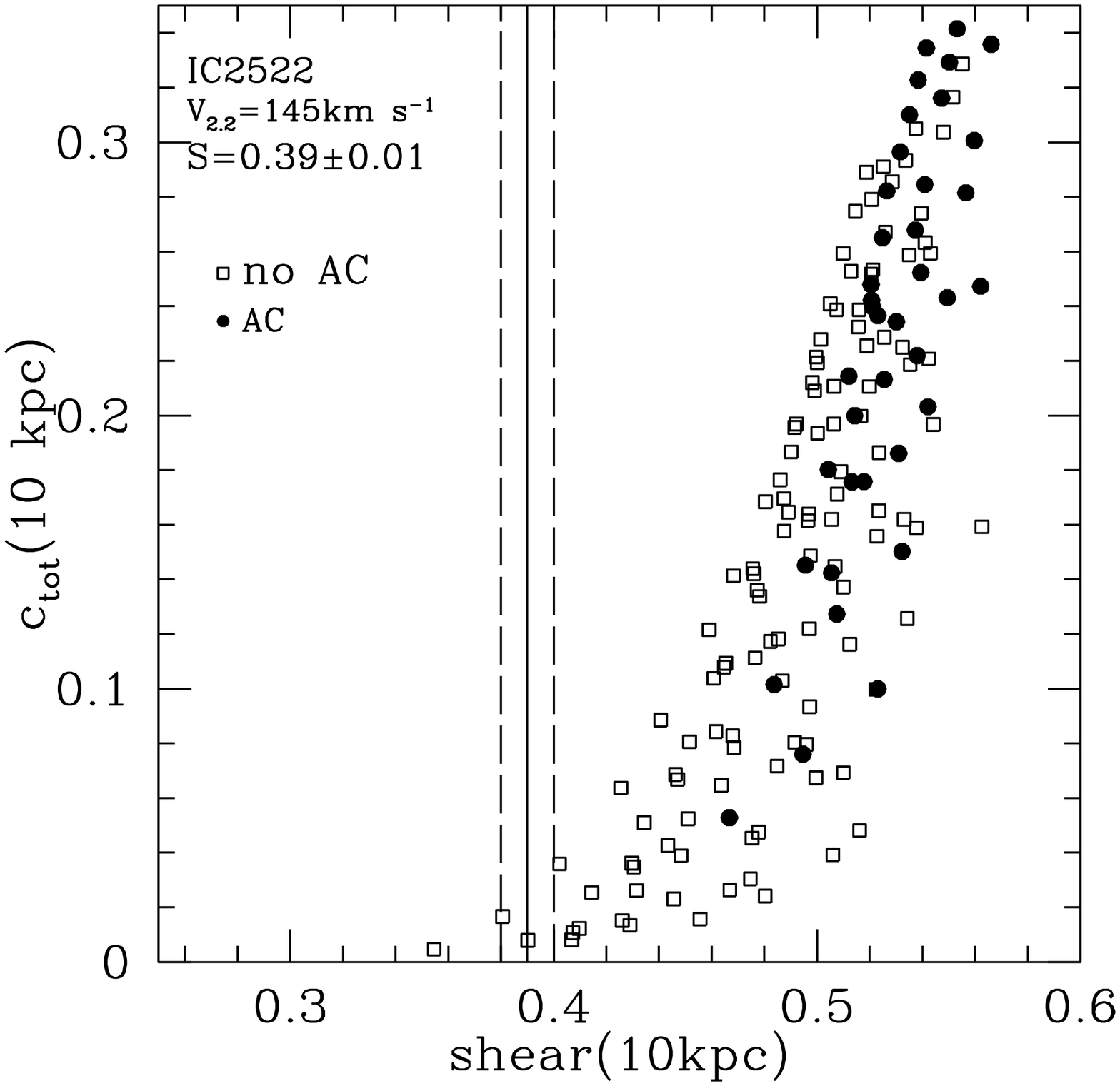}{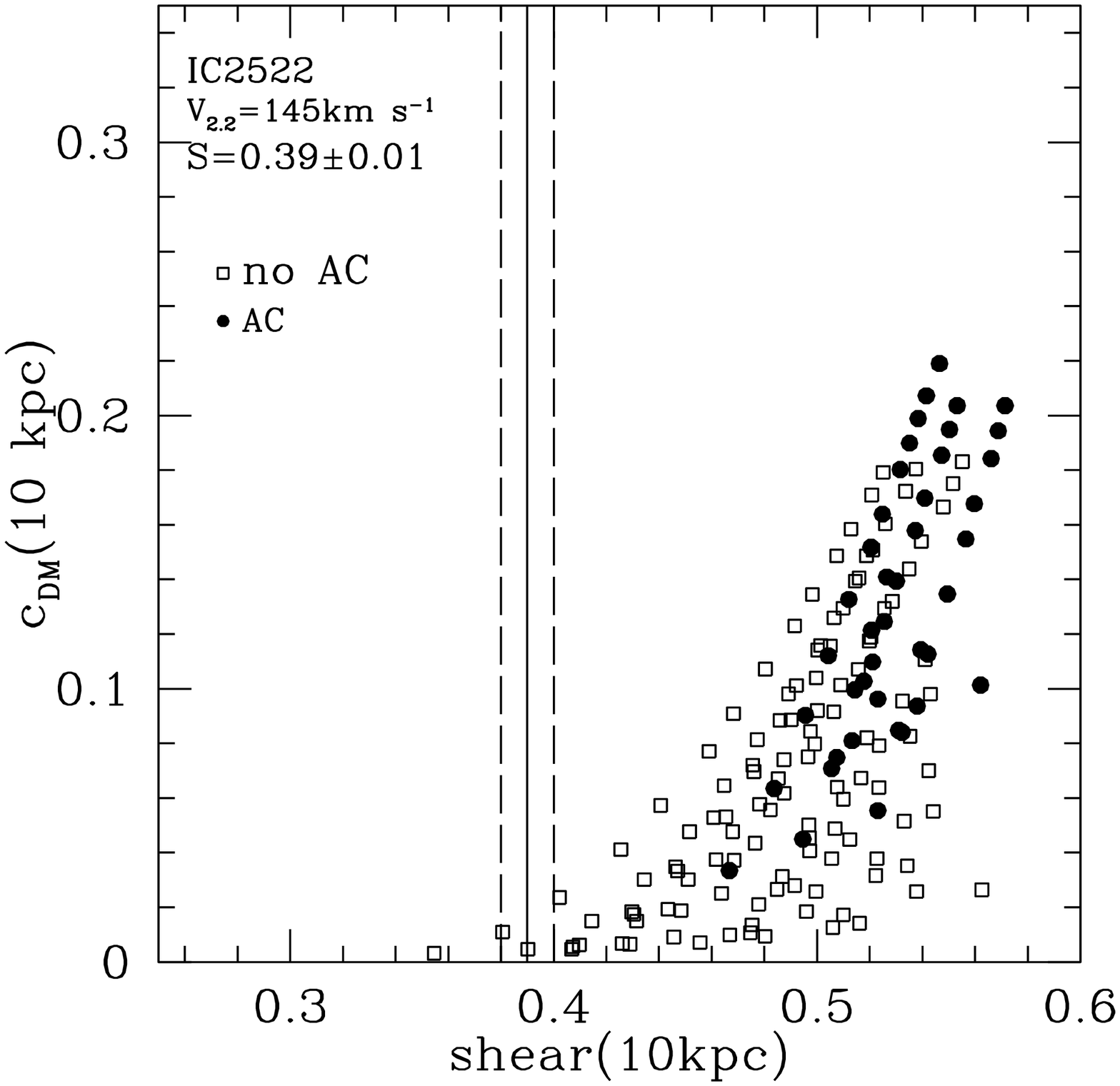}
\caption{Model results for IC 2522. {\em Left}: Central mass 
concentration (mass fraction within $10$ kpc) versus shear, {\em right}:
dark matter concentration (dark matter mass fraction within $10$ kpc)
versus shear. The open squares represent the model without
adiabatic contraction of the stellar halo
and solid circles represnt the model with adiabatic
contraction. 
The vertical lines present the measured shear (solid line) with the 1-$\sigma$
error added and subtracted (dashed lines).}
\label{ic2522}
\end{figure*}

Model results for the central mass concentration in each galaxy
are shown in Figures \ref{eso582} and \ref{ic2522}.
Here $c_{\rm tot}$ and $c_{\rm dm}$ correspond to the fraction of
the total and dark matter mass contained within central
$10$ kpc of each system.  We see that ESO 582-G12 contains roughly
$20$\% of its mass within $10$ kpc, while IC 2552 is quite diffuse
with $c_{\rm tot} < 5$\%.

%JB: rearranged and modified...

What drives these differences?  Take
IC 2522 for example, which 
has a very low concentration. This arises because of the
combination of its (small) disk scalelength {\em and} the fact
that the rotation curve is rising at 10 kpc (implied by a shear
of 0.39$\pm$0.1). The rotation curve of the stellar disk alone
peaks at 2.2 scalelengths ($\sim 8.8$ kpc), and will be falling
at 10 kpc (where the shear is measured). The measured shear
of 0.39$\pm$0.01 at 10 kpc implies that the total rotation curve
is rising at 10 kpc, which means the dark matter must be very
extended (or of low concentration).
ESO 582-G12, on the other hand, has a falling rotation curve at
10 kpc, but a {\em larger} disk scalelength. This requires a
more concentrated halo to explain the observed shear rate.

The two galaxies chosen for this modeling are, in many ways, quite 
comparable. Their Hubble types, B/D ratios and disk luminosities are
very similar. The only significant difference is in their measured
rates of shear. The differences in their favored formation models
(AC for ESO 582-G12 versus non-AC for IC 2522), halo masses and
concentrations, comes down to a different measurement for their
shear and the sizes of their disks.  It is quite interesting that
two galaxies, which seem so similar in size and Hubble type,
seem to inhabit very different types of dark matter halos.
It is tempting to speculate that the shear rate itself 
provides an underlying physical driver to push galaxies in 
drastically different dark matter halos towards similar luminosities.
However, the problem would then be what stops such a drive and makes the 
galaxy luminosities so similar, when the halo masses show large differences.

%%%% I added this paragraph based on your reply to Aaron's comments

%%%%

\begin{figure*}
\plottwo{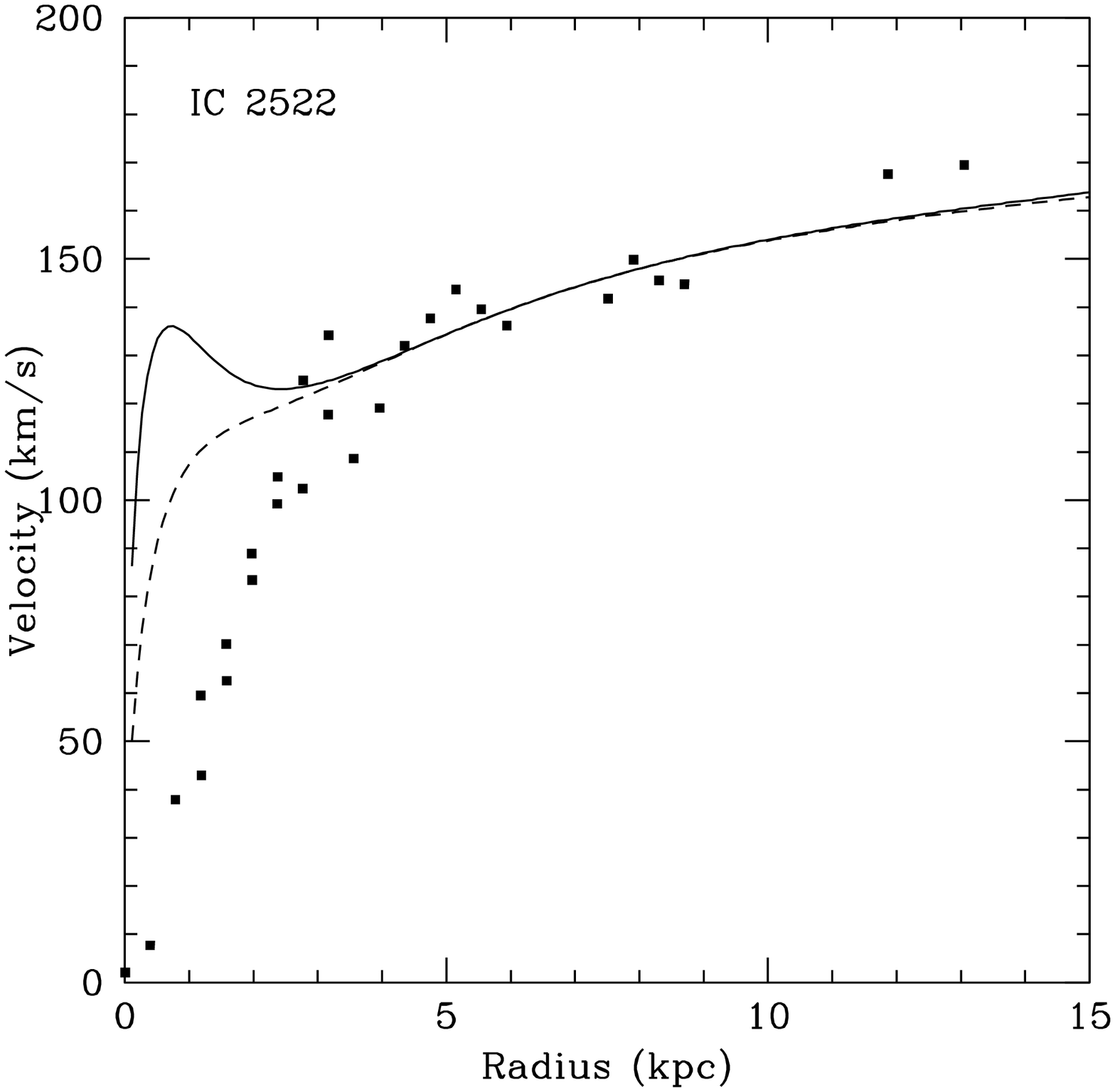}{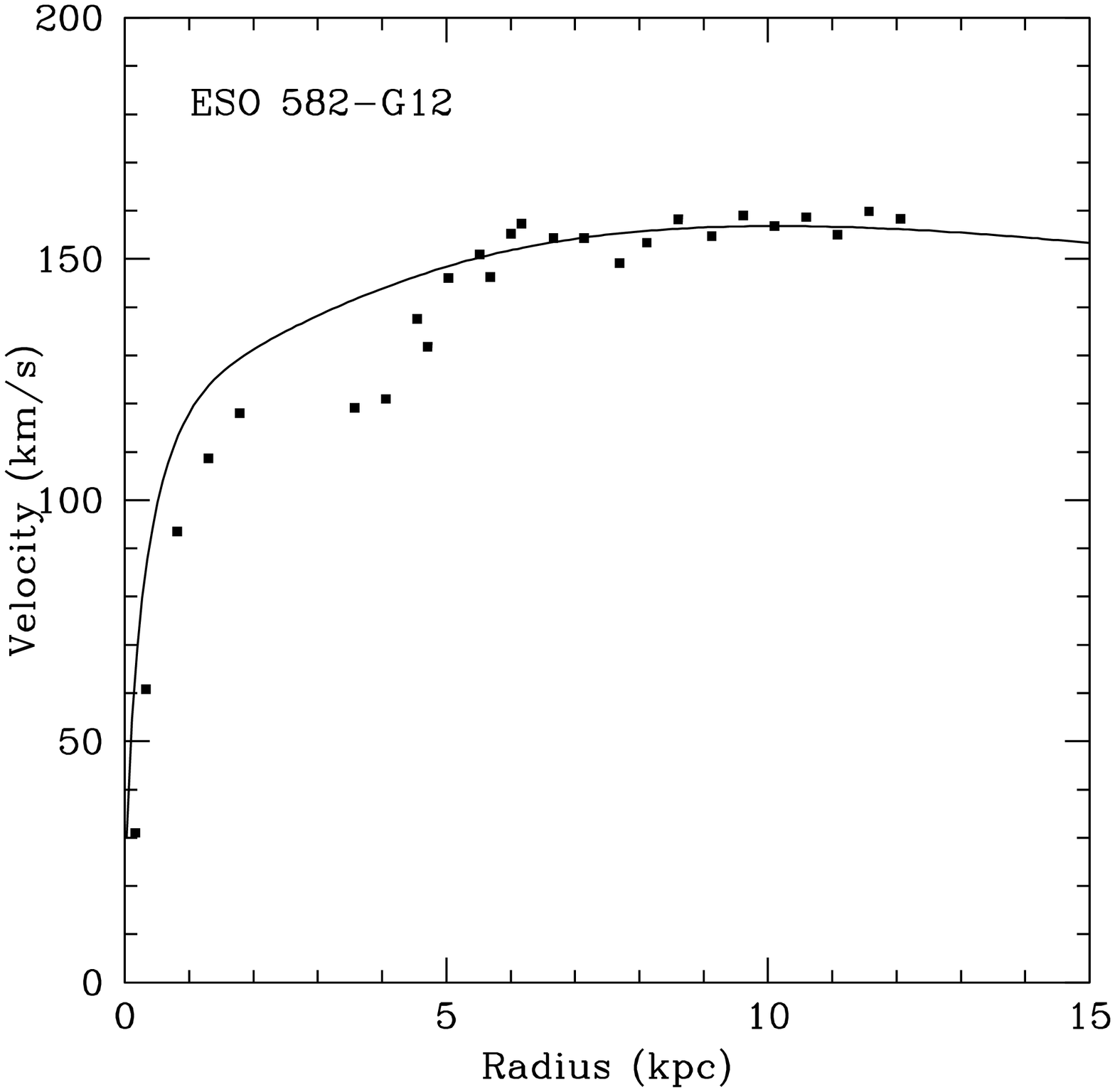}
\caption{Observed rotation curves with the overlaid model rotation curves for 
{\em Left}: IC 2522 and {\em Right}: ESO 582-G12. The errors on the data points
are typically $<$10\% (Persic \& Salucci 1995). The solid lines represent the
rotation curve that best matches the observed shear at 10 kpc. For IC 2522
the dashed line represents a model rotation curve with a bulge that has
an affective radius which is a factor of 2 larger than measured in the
1-D decomposition, in order to mimic the affect of a bar with a major-to-minor
axis ratio of $\sim$2.}
\label{rotation}
\end{figure*}

This clearly demonstrates that the  shear  rate   adds  an
important  constraint on galaxy formation models  compared to what can
be learned from standard  Tully-Fisher constraints alone.  Indeed, the
results for IC 2522 and ESO 582-G12 are  tantalizing.  The first, with
a low shear  rate, favors a  non-adiabatically contracted halo, with a
low halo concentration and a massive, extended dark  matter halo.  The
second, with a high shear rate, favors an adiabatically contracted halo
with a high NFW concentration, and  relatively low virial mass.  These
two cases motivate the application  of the shear-rate constraint to  a
larger sample of  galaxies.  Since shear is  related to the spiral arm
pitch angle and spiral arms are clearly detectable in disk galaxies at
$z\sim 1$ (e.g. Elmegreen, Elmgreen \& Hirst  2004), it is possible to
constrain models of disk formation and dark matter halo structure back
to a look-back time   of $\sim 7$  Gyrs, and   evaluate how  the  mass
distribution in disk galaxies changes with time.

\begin{figure}
\vspace*{3in}
%\plotone{seigar_f10.eps}
\caption{$K_s$ band image of IC 2522, revealing a bar that was not eveident
in the optical imaging of this galaxy. The image is aligned such that North
is up and the field is 3.3$\times$3.3 arcmin.}
\label{ic2522_k}
\end{figure}

An important test of our modeling is how well the modeled rotation 
curves fit the data. Figure 9 shows the observed rotation curves
from Persic \& Salucci (1995) overlaid with our modeled rotation curves.
In both cases the rotation curve is modeled well in the outer parts,
where we are measuring the shear. In the inner regions, where the
bulge is important, both rotation curve models overestimate the
observed rotation curve, although the model for ESO 582-G12 presents
a much better case than that for IC 2522. For IC 2522, the solid line
represents a model rotation curve, with a bulge-to-disk ratio of 0.164
(see Table 4) and a bulge effective radius of 0.2 kpc (as measured in the
1-D decomposition). However, infrared imaging of IC 2522 (see Figure 10)
reveals that this galaxy has a bar with a major axis length $\sim$2 kpc
(i.e.\ the radius within which the model overestimates the
observed rotational velocities) and a major-to-minor axis ratio of 
$\sim$2. (It is common for near-IR imaging to 
reveal bars that may have been hidden due to dust extinction in an optical
image, e.g.\ Seigar \& James 1998a; Eskridge et al. 2000; Seigar 2002; 
Seigar et al. 2003).
The spectroscopic data taken by Mathewson et  al.\ (1992) were observed with 
a slit aligned along the major axis of each galaxy, and in the case
of IC 2522, the major axis of the bar is well aligned with the galaxy 
major axis. The stellar orbits within a bar are such that the dominant
motion is parallel to its major-axis (e.g.\ Athanassoula 1992), 
and so this will account (to some
extent) for the low values of the measured rotation velocities (compared to 
the modeled values) within a 
few kpc. Furthermore, if we double the effective radius of the modeled
bulge, the resulting rotation curve is the dashed line in Figure 9. This still
fits the outer part of the galaxy rotation curve extremely well, and
the central velocities are not as badly overestimated. Modifying the mass
models in this way does not affect the trends displayed in Figures 6 and 8,
it just affects how well the model reproduces the inner part of the
observed rotation curve.
This is probably because the main constraints we are using to model these
galaxies are at a 10 kpc radius, well outside the region where the bulge 
or bar is
dominating the rotational velocities. As a result, the concentrations we
determine from our modeling do not depend on the mass distribution assumed
for the bulge (i.e.\ the mass distribution within the central few kpc).
This approach is, therefore, a powerful and robust method, because its
constraints are relatively insensitive to the details of bars versus
bulges in the central regions.

\section{Conclusions}

In this paper we have shown that near-infrared and optical spiral arm 
pitch angles are the same, on average, and as a result we have strengthened
the correlation between spiral arm pitch angle and shear rate originally
shown in Seigar et al.\ (2005) and expanded here using optical data.
Using an infall model, we have shown that the use of rates of shear
(which can now be derived from spiral arm pitch angles) allow us
to put constraints on the total central mass concentration, the
dark matter concentration and the initial NFW concentration. In some
cases it may be possible to determine if the infall has to occur 
adiabatically or non-adiabatically, and this is demonstrated by
IC 2522 which has to undergo non-adiabatic infall.

This method can be used to determine the central concentrations
of galaxies and to constrain galaxy formation models in any galaxy 
that has detectable spiral structure. In future papers we will apply
our technique to a large sample of galaxies. Furthermore, since
spiral structure can clearly be seen
in disk galaxies out to $z\sim 1$ (Elmegreen et al.\ 2004), the evolution
of central mass concentrations in disk galaxies can be estimated as a 
function of look-back time, and we plan to investigate this as well.

\acknowledgments

Support for this work was provided by NASA through grant number 
HST-AR-10685.01-A
from the Space Telescope Science Institute, which is operated by the
Association of Universities for Research in Astronomy, Inc., under NASA
contract NAS5-26555. 
This research has made use of the NASA/IPAC Extragalactic Database (NED) 
which is operated by the Jet Propulsion Laboratory, California Institute 
of Technology, under contract with the National Aeronautics and Space 
Administration. This work also  made use of data from the Ohio State 
University Bright Spiral Galaxy Survey, which was funded by grants 
AST-9217716 and AST-9617006 from the United States National Science 
Foundation, with additional support from the Ohio State University.
The authors wish to thank Paolo Salucci for supplying the rotation curve
data from which shear and the $V_{2.2}$ velocities were derived.
We also wish to show our appreciation to the referee, Dr.\ P.\ Grosbol,
for his many useful comments.

\end{document}